\documentclass[fleqn,usenatbib]{mnras}

\usepackage{newtxtext,newtxmath}

\usepackage[T1]{fontenc}

\DeclareRobustCommand{\VAN}[3]{#2}
\let\VANthebibliography\thebibliography
\def\thebibliography{\DeclareRobustCommand{\VAN}[3]{##3}\VANthebibliography}

\usepackage{graphicx}
\usepackage{amsmath}
\usepackage{caption}

\title[UV continuum slopes at $z \simeq 8 - 16$]{\vspace{-0.2cm}The ultraviolet continuum slopes ($\mathbf{\beta}$) of galaxies at $\mathbf{z\simeq8-16}$ from JWST and ground-based near-infrared imaging\vspace{-0.2cm}}

\author[F. Cullen et al.]{Fergus Cullen $^{1}$\thanks{E-mail: fc@roe.ac.uk},
R.\,J. McLure$^{1}$,
D.\,J. McLeod$^{1}$,
J.\,S. Dunlop$^{1}$,
C.\,T. Donnan$^{1}$,
A.\,C. Carnall$^{1}$,
\newauthor
R.\,A.\,A. Bowler$^{2}$,
R. Begley$^{1}$,
M.\,L. Hamadouche$^{1}$,
T.\,M. Stanton$^{1}$
\\
$^{1}$Institute for Astronomy, University of Edinburgh, Royal Observatory, Edinburgh, EH9 3HJ, UK\\
$^{2}$Jodrell Bank Centre for Astrophysics, University of Manchester, Oxford Road, Manchester, M13 9PL, UK\vspace{-0.3cm}
}

\date{Accepted XXX. Received YYY; in original form ZZZ\vspace{-0.2cm}}

\pubyear{2022}

\begin{document}
\label{firstpage}
\pagerange{\pageref{firstpage}--\pageref{lastpage}}
\maketitle

\begin{abstract}
We study the rest-frame ultraviolet (UV) continuum slopes ($\beta$) of galaxies at redshifts $8 < z < 16$ ($\langle z \rangle = 10$), using a combination of {\it James Webb Space Telescope} (JWST) ERO and ERS NIRCam imaging and ground-based near-infrared imaging of the COSMOS field. 
The combination of JWST and ground-based imaging provides a wide baseline in both redshift and absolute UV magnitude ($-22.6 < M_{\rm UV} < -17.9$), sufficient to allow a meaningful comparison to previous results at lower redshift. 
Using a power-law fitting technique, we find that our full sample (median $M_{\rm UV}=-19.3\pm 1.3$) returns an inverse-variance weighted mean value of $\langle \beta \rangle = -2.10 \pm 0.05$, with a corresponding median value of $\beta=-2.29 \pm 0.09$.
These values imply that the UV colours of galaxies at $z>8$ are, \emph{on average}, no bluer than the bluest galaxies in the local Universe (e.g., NGC 1705; $\beta=-2.46$).
We find evidence for a $\beta-M_{\rm UV}$ relation, such that brighter UV galaxies display redder UV slopes ($\rm{d}\beta/ \rm{d} M_{\rm UV} = -0.17 \pm 0.05$). 
Comparing to results at lower redshift, we find that the slope of our $\beta-M_{\rm UV}$ relation is consistent with the slope observed at $z\simeq 5$ and that, at a given $M_{\rm UV}$, our $8<z<16$ galaxies are bluer than their $z\simeq 5$ counterparts, with an inverse-variance weighted mean offset of $\langle \Delta \beta \rangle = -0.38 \pm 0.09$.
We do not find strong evidence that any objects in our sample display ultra-blue UV continuum slopes (i.e., $\beta\lesssim-3$) that would require their UV emission to be dominated by ultra-young, dust-free stellar populations with high Lyman-continuum escape fractions.
Comparing our results to the predictions of theoretical galaxy formation models, we find that the galaxies in our sample are consistent with the young, metal-poor and moderately dust-reddened galaxies expected at $z>8$.
\end{abstract}

\begin{keywords}
galaxies: evolution - galaxies: formation - galaxies: high-redshift - galaxies: starburst -  dark ages, reionization, first stars
\vspace{-0.5cm}
\end{keywords}



\section{Introduction}

Constraining the physical properties of the first galaxies is a key goal of the \emph{James Webb Space Telescope} (JWST).
Within the first month of the data being released, JWST has already revealed a substantial population of previously unseen galaxies at $z>10$ \citep[e.g.,][]{adams2022,atek2022,castellano2022,naidu2022,donnan2022, finkelstein2022, harikane2022}, as well as one galaxy candidate at $z\simeq16$ \citep[within $\approx 200$ Myr of the Big Bang;][]{donnan2022}.
These galaxies provide an unprecedented opportunity to study the properties of primordial stellar populations in the early Universe.

One potential indicator of ultra-young, ultra-low metallicity, stellar populations is the power-law index of the rest-frame ultraviolet (UV) continuum, $\beta$, where $f_{\lambda} \propto \lambda^{\beta}$.
At very young ages and low metallicities (e.g., $t < 30$ Myr and $Z_{\star}\lesssim10^{-3}$), and in the absence of dust extinction and nebular continuum emission, a very low (i.e., blue) value of $\beta \simeq -3$ is expected \citep[e.g.,][]{schaerer2002, bouwens2010,chisholm2022}.
A robust determination of $\beta=-3$ would unequivocally indicate a stellar population that has recently formed from pristine (or near-pristine) gas with a large ionizing photon escape fraction \citep[e.g.,][]{robertson2010}.
Finding such galaxies would therefore have important implications for our understanding of the first galaxies and the process of cosmic hydrogen reionization.

In the pre-JWST era, no strong evidence for such primordial ${\beta=-3}$ populations was found.
Studies of faint galaxies up to $z\simeq 7-8$ with the \emph{Hubble Space Telescope} (HST) revealed an average power-law index of ${\langle \beta \rangle \simeq -2}$, indicating moderately young and metal-poor, but in no sense extreme, stellar populations \citep[e.g.,][]{dunlop2013, finkelstein2012, bouwens2014}.
Indeed, $\beta \simeq -2$ to $\beta \simeq -2.5$ is typical of the bluest galaxies observed at $z = 2-4$ \citep[e.g.,][]{mclure2018}, and even in the local Universe \citep[e.g., NGC 1705, $\beta=-2.46\pm0.01$, $M_{\rm UV} = -18$;][]{calzetti1994, vazquez2004}.
Early claims of extremely blue (i.e., $\beta \leq -3$) galaxies from HST imaging were later shown to be the result of an observational bias, pushing measurements towards artificially blue $\beta$ values for faint sources near the detection threshold \citep{bouwens2010, dunlop2012, rogers2013}.

By providing unprecedentedly deep infrared imaging up to ${\lambda = 5\, \mu \rm{m}}$, JWST/NIRCam now enables the first robust estimates of $\beta$ for galaxies at $z > 8$.
Recently, \cite{topping2022} have provided the first measurements of $\beta$ at $z\simeq7-11$ from JWST/NIRCam imaging in the EGS field.
They find a median value of $\beta=-2.0$, consistent with the blue, but otherwise unremarkable, populations found with \emph{HST}.
Interestingly, however, \cite{topping2022} also report two galaxies with seemingly secure $\beta \simeq -3$ measurements.
The ultra-blue UV slopes inferred for these sources are bolstered by a lack of strong nebular emission-line signatures in the rest-frame optical photometry, indicating the large ionizing photon escape fractions expected for such a population.

In this paper we use the new galaxy sample described in \cite{donnan2022} to present a complementary study of UV continuum slopes for $\rm{N}=61$ galaxies in the redshift range $z~\simeq~8-16$ (with mean $\langle z \rangle \simeq 10$). 
The \cite{donnan2022} sample combines galaxies drawn from the early JWST deep fields (at $z\simeq9-16$) with an additional sample selected from wide-area ground-based near-IR imaging in the COSMOS/UltraVISTA field (at $z\simeq8-10$).
Crucially, with the inclusion of this ground-based sample, we can also probe the brightest galaxies at these redshifts, which evade current JWST surveys. 
This extended baseline in UV luminosity enables us to investigate $\beta$ across a factor of $\simeq 80$ in UV luminosity, placing early constraints on the relationship between $\beta$ and UV magnitude at $z>8$ \citep[i.e., the $\beta-M_{\rm UV}$, or the colour-magnitude, relation;][]{rogers2014}.

Our aim is to provide an exploratory study of the constraints on $\beta$ at $z>8$ enabled by deep JWST multi-band imaging, and to critically assess any early evidence for an evolution in the typical $\beta$ values, as well as the relation between $\beta$ and $M_{\rm UV}$ at these redshifts.
We also examine evidence for any robust $\beta \simeq -3$ sources in our sample, and discuss the possibility of spurious $\beta \leq -3$ detections for faint sources in the new JWST imaging.

The paper is structured as follows.
In Section \ref{section:data} we describe the data and galaxy sample constructed by \cite{donnan2022}, and provide the details of our method for determining $\beta$.
In Section \ref{section:results} we present our $\beta$ measurements and outline the main results of our analysis.
In Section \ref{section:discussion} we discuss the implications for our results before summarising our main conclusions in Section \ref{section:conclusion}.
Throughout we use the AB magnitude system \citep{oke1974,oke1983}, and assume a standard cosmological model with $H_0=70$\,km s$^{-1}$ Mpc$^{-1}$, $\Omega_m=0.3$ and $\Omega_{\Lambda}=0.7$.

\section{Data and UV continuum slope fitting}\label{section:data}

\subsection{JWST NIRCam imaging}

Our JWST sample was initially presented in \citet{donnan2022}. 
The sample is drawn from public NIRCam imaging of three fields (SMACS J0723, GLASS and CEERS) released as part of the Early Release Observations (ERO, see \citealt{Pontoppidan2022}) and Early Release Science (ERS) programmes \citep{treu2022}.
Each of the three JWST fields were imaged in a combination of the F090W, F115W, F150W, F200W, F277W, F356W, F410M and F444W filters, with the specific combination of filters varying slightly from field to field \citep[see Table 2 of][]{donnan2022}.
This JWST NIRCam imaging was reduced using \texttt{PENCIL} (PRIMER enhanced NIRCam Image Processing Library) which is a custom version of the JWST pipeline (1.6.2) with additional steps for background subtraction and the removal of `snowball' artefacts and including up-to-date calibrations and zero-point corrections \citep[see][]{donnan2022}.
The final combined JWST NIRCam imaging area totalled $\simeq 45$\,arcmin$^2$ (with some variation between filters).
For this work, prior to catalogue construction, all of the NIRCam imaging was homogenized to the point-spread-function (PSF) of the F444W filter.
\begin{table}
    \centering
    \caption{The best-fitting UV continuum slopes ($\beta$) for the full sample of galaxies at $z>7.5$ in the combined JWST and COSMOS/UltraVISTA samples.
    The first column gives the source ID taken from \citet{donnan2022}. 
    Column two gives the sample (COSMOS/UltraVISTA or JWST).
    Columns three and four give the photometric redshift ($z_{\rm phot}$) and absolute UV magnitude ($M_{\rm UV}$) taken from \citet{donnan2022}.
    Column four gives the derived UV continuum slope $\beta$.}
    \label{tab:sample}
    \renewcommand{\arraystretch}{1.35} 
    \begin{tabular}{lcccc} 
        \hline
        ID & Sample & $z_{\rm phot}$ & $M_{\rm UV}$ & $\beta$ \\
        \hline
        334330 & COSMOS & $7.58$ & $-21.30$ & $-1.01^{+0.62}_{-0.64}$\\
        733875 & COSMOS & $7.58$ & $-21.57$ & $-1.87^{+0.74}_{-0.79}$\\
        812867 & COSMOS & $7.58$ & $-21.02$ & $-2.19^{+0.62}_{-0.72}$\\
        688541 & COSMOS & $7.66$ & $-22.15$ & $-2.88^{+0.38}_{-0.41}$\\
        765906 & COSMOS & $7.66$ & $-22.61$ & $-1.26^{+0.27}_{-0.25}$\\
        626972 & COSMOS & $7.75$ & $-21.49$ & $-2.65^{+0.71}_{-1.03}$\\
        536767 & COSMOS & $8.02$ & $-21.40$ & $-2.16^{+0.84}_{-0.88}$\\
        861605 & COSMOS & $8.02$ & $-21.33$ & $-3.91^{+1.10}_{-1.60}$\\
        978389 & COSMOS & $8.02$ & $-21.68$ & $-1.95^{+0.75}_{-1.12}$\\
        484075 & COSMOS & $8.11$ & $-22.05$ & $-2.07^{+0.48}_{-0.50}$\\
        578163 & COSMOS & $8.20$ & $-22.35$ & $-1.04^{+0.31}_{-0.29}$\\
        458445 & COSMOS & $8.38$ & $-21.65$ & $-2.89^{+0.88}_{-1.11}$\\
        448864 & COSMOS & $8.57$ & $-21.15$ & $-2.29^{+0.67}_{-0.81}$\\
        306122 & COSMOS & $8.76$ & $-21.76$ & $-2.25^{+0.49}_{-0.53}$\\
        892014 & COSMOS & $8.96$ & $-22.16$ & $-1.97^{+0.39}_{-0.45}$\\
        817482 & COSMOS & $9.89$ & $-22.57$ & $-1.15^{+0.56}_{-0.59}$\\
        43031 & JWST & $8.57$ & $-18.43$ & $-2.17^{+0.39}_{-0.41}$\\
        29274$\_$4 & JWST & $8.86$ & $-18.41$ & $-2.55^{+1.28}_{-1.15}$\\
        1434$\_$2 & JWST & $9.16$ & $-18.82$ & $-2.28^{+0.45}_{-0.45}$\\
        44085 & JWST & $9.26$ & $-18.25$ & $-1.41^{+0.41}_{-0.38}$\\
        38697 & JWST & $9.36$ & $-18.86$ & $-1.77^{+0.50}_{-0.46}$\\
        5071 & JWST & $9.47$ & $-18.02$ & $-2.55^{+1.00}_{-1.01}$\\
        44711 & JWST & $9.47$ & $-20.14$ & $-2.12^{+0.16}_{-0.15}$\\
        43866 & JWST & $9.47$ & $-18.14$ & $-2.78^{+0.32}_{-0.31}$\\
        34086 & JWST & $9.47$ & $-17.87$ & $-1.92^{+0.25}_{-0.23}$\\
        14391 & JWST & $9.47$ & $-18.81$ & $-1.80^{+0.51}_{-0.44}$\\
        12682 & JWST & $9.57$ & $-18.95$ & $-1.38^{+0.50}_{-0.56}$\\
        44566 & JWST & $9.68$ & $-20.68$ & $-1.64^{+0.11}_{-0.12}$\\
        22480 & JWST & $9.68$ & $-18.50$ & $-1.60^{+0.58}_{-0.52}$\\
        15019 & JWST & $9.68$ & $-18.67$ & $-4.61^{+1.16}_{-1.96}$\\
        12218 & JWST & $9.68$ & $-19.28$ & $-2.47^{+0.27}_{-0.28}$\\
        3398 & JWST & $9.68$ & $-18.21$ & $-5.72^{+1.37}_{-2.18}$\\
        6200 & JWST & $9.79$ & $-18.52$ & $-3.03^{+0.88}_{-1.27}$\\
        7606 & JWST & $9.89$ & $-18.08$ & $-4.28^{+2.21}_{-2.90}$\\
        3763 & JWST & $9.89$ & $-18.99$ & $-3.41^{+0.39}_{-0.38}$\\
        1698 & JWST & $10.45$ & $-20.62$ & $-2.00^{+0.14}_{-0.15}$\\
        20976$\_$4 & JWST & $10.45$ & $-18.80$ & $-1.53^{+0.58}_{-0.61}$\\
        6647 & JWST & $10.45$ & $-18.88$ & $-0.23^{+1.12}_{-0.92}$\\
        3710 & JWST & $10.45$ & $-19.06$ & $-2.05^{+0.54}_{-0.53}$\\
        4063 & JWST & $10.45$ & $-18.03$ & $-3.13^{+0.74}_{-0.77}$\\
        30585 & JWST & $10.56$ & $-19.35$ & $-2.98^{+0.38}_{-0.39}$\\
        \hline
    \end{tabular}
\end{table}

\begin{table}
\ContinuedFloat
    \centering
    \caption{Continued.}
    \label{tab:sample1}
    \def\arraystretch{1.35}
    \begin{tabular}{lcccc} 
        \hline
        ID & Sample & $z_{\rm phot}$ & $M_{\rm UV}$ & $\beta$ \\
        \hline
        73150 & JWST & $10.56$ & $-19.07$ & $-3.57^{+0.95}_{-1.00}$\\
        21071$\_$2 & JWST & $10.68$ & $-19.27$ & $-2.71^{+0.64}_{-0.58}$\\
        20757 & JWST & $10.68$ & $-17.88$ & $-0.74^{+1.02}_{-1.05}$\\
        6415 & JWST & $10.79$ & $-19.13$ & $-2.02^{+1.02}_{-1.05}$\\
        120880 & JWST & $10.79$ & $-19.43$ & $-2.73^{+0.58}_{-0.59}$\\
        26598 & JWST & $10.79$ & $-18.47$ & $-3.31^{+0.80}_{-0.87}$\\
        61486 & JWST & $11.15$ & $-19.61$ & $-2.61^{+0.41}_{-0.52}$\\
        622$\_$4 & JWST & $11.27$ & $-18.92$ & $-3.38^{+0.56}_{-0.58}$\\
        33593$\_$2 & JWST & $11.27$ & $-19.58$ & $-2.07^{+0.28}_{-0.30}$\\
        77241 & JWST & $11.27$ & $-19.60$ & $-2.51^{+0.38}_{-0.42}$\\
        5268$\_$2 & JWST & $11.40$ & $-19.16$ & $-2.41^{+0.52}_{-0.55}$\\
        127682 & JWST & $11.40$ & $-19.07$ & $-2.73^{+0.60}_{-0.64}$\\
        26409$\_$4 & JWST & $11.90$ & $-18.84$ & $-3.25^{+1.02}_{-1.47}$\\
        8347 & JWST & $11.90$ & $-19.09$ & $-2.93^{+0.33}_{-0.38}$\\
        10566 & JWST & $12.03$ & $-19.70$ & $-3.44^{+0.43}_{-0.43}$\\
        32395$\_$2 & JWST & $12.29$ & $-19.89$ & $-3.30^{+0.25}_{-0.30}$\\
        1566 & JWST & $12.29$ & $-18.77$ & $-2.51^{+0.51}_{-0.55}$\\
        17487 & JWST & $12.42$ & $-20.89$ & $-2.64^{+0.26}_{-0.27}$\\
        27535$\_$4 & JWST & $12.56$ & $-19.42$ & $-1.70^{+0.44}_{-0.46}$\\
        93316 & JWST & $16.39$ & $-21.66$ & $-1.89^{+0.15}_{-0.15}$\\
    \hline
    \end{tabular}
\end{table}

The JWST catalogues were created by running \textsc{Source Extractor} \citep{Bertin1996} in dual-image mode.
The F200W image was used as the detection image to optimise for the selection $z\geq8$ galaxies.
The photometry for each JWST target was computed in both 0.5-arcsec and 0.36-arcsec diameter apertures.
For the purposes of the present paper, we adopt the 0.5-arcsec apertures to prevent biases in $\beta$ measurements in more extended sources \citep{rogers2014}.
However, we have confirmed that adopting the 0.36-arcsec diameter apertures would not change our main results.
Redshifts for each object were estimated using the photometric redshift fitting code \texttt{EAZY} \citep{brammer2008}.
A thorough selection procedure, described in \cite{donnan2022}, resulted in a final sample of $45$ galaxies at $z>8.5$ across the three fields.

Absolute rest-frame UV magnitudes ($M_{\rm UV}$) were calculated for each object by integrating the best-fitting \texttt{EAZY} spectral energy distribution (SED) through a tophat filter centered on $\lambda_{\rm rest}=1500\, \text{\AA}$ \citep{donnan2022}.

\subsection{COSMOS UltraVISTA}
Our COSMOS sample was also initially presented in \citet{donnan2022}.
The sample was drawn from the UltraVISTA survey \citep{mccracken2012} which provides deep $YJHK_s$ near-IR imaging across 1.8\,deg$^2$ in the COSMOS field.
The deep near-IR imaging is supplemented with optical imaging in $u^*griz$ from the CFHT Legacy Survey \citep{hudelot2012}, and the $GRIZy$+NB816+NB921 filters from the Hyper Suprime-Cam Subaru Strategic Program (HSC-SSP) DR2 \citep{aihara2019}.
All of the near-infrared and optical imaging in COSMOS was aligned to the GAIA EDR3 reference and PSF-homogenised to the UltraVISTA $Y-$band.
Additionally, the COSMOS/UltraVISTA dataset was further augmented by $3.6\mu$m and $4.5\mu$m photometry from \textit{Spitzer}/IRAC imaging provided by the Cosmic Dawn Survey \citep{moneti2022}.

The COSMOS catalogue was produced from inverse variance weighted stacks of the data in the $Y$, $J$, $H$, and $K_s-$bands as described in \cite{donnan2022}.
Photometric redshifts were estimated with \texttt{EAZY} and, after applying a robust selection criteria, we retained a final sample of 16 galaxies at $z>7.5$.
Absolute rest-frame UV magnitudes were calculated for each object from the best-fitting \texttt{EAZY} SED.
Combined, our JWST and COSMOS/UltraVISTA samples yielded a total of 61 galaxies at $z\simeq8-16$.

\subsection{Measuring the UV continuum slope}\label{subsec:beta_fitting}

A number approaches to determining the UV continuum slope from broadband photometry have been presented in the literature, including single colour measurements \citep[e.g.,][]{mclure2011, dunlop2012, dunlop2013} and SED template fitting \citep[e.g.,][]{finkelstein2012, tacchella2022}.
Here we have adopted the power-law fitting method advocated by \cite{rogers2014} in their study of the $\beta-M_{\rm UV}$ relation at $z\simeq5$.
For each source, the redshift was fixed to the best-fitting photometric redshift estimated by \cite{donnan2022} and the photometry covering rest-frame wavelengths $\lambda_{\rm rest} \leq 3000\, \text{\AA}$ was modelled as a pure power law ($f_{\lambda} \propto \lambda^{\beta}$), with IGM absorption at $\lambda \leq 1216\, \text{\AA}$ included using the \cite{inoue2014} prescription.
The only free parameter in this approach is $\beta$, the power-law spectral index of the UV continuum red-ward of $\lambda = 1216\, \text{\AA}$.
We allowed $\beta$ to vary over the range $-10 \leq \beta \leq 10$ and used the nested sampling code \texttt{dynesty} \citep{speagle2020} to sample the full posterior distribution assuming a uniform prior.
The derived values of $\beta$ for our full sample are given in Table \ref{tab:sample}.

We investigated the effect of redshift uncertainties by running an additional set of fits in which redshift ($z$) was included as a free parameter.
For the prior on redshift we assumed a Gaussian centered on the best-fitting photometric redshift from \citet{donnan2022} ($z_{\rm phot}$) with ${\sigma_{z}=1}$.
We find that the effect of fitting for redshift on the derived values of $\beta$ is negligible, with a median difference across the sample of $\Delta \beta = 0.04$, corresponding to a median difference in redshift of $(z-z_{\rm phot})=-0.03$.
However, we find that marginalizing over a plausible range of redshifts in this way increases the typical error on $\beta$ by $\simeq 13\%$.
For the the purposes of this paper, we decided to fix our redshifts to the more-accurate $z_{\rm phot}$ values presented in \citet{donnan2022} (i.e., which are derived from fitting to the full rest-frame UV to optical photometry) but increased the corresponding $\beta$ uncertainties by a factor $1.13$.

We note that our approach is similar to the power-law fitting method used by \cite{topping2022}, with the main difference being that our IGM model enables us to include filters encompassing the Lyman break.
However, our results are unchanged if we restrict the fitting to filters red-ward of $1216 \, \text{\AA}$.
Finally, it is also worth noting that we have explicitly assumed that any emission lines present in the UV spectrum - in particular $\mathrm{Ly}\alpha$ - have a negligible impact on the observed photometry.
This assumption appears to be justified based on the early, low-resolution, JWST/NIRSpec spectroscopy of $z>9$ sources \citep[e.g.][]{ curtis-lake2022,roberts-borsani2022}.

\section{Results}\label{section:results}

\begin{figure}
        \vspace*{-0.3in}
        \centerline{\includegraphics[width=\columnwidth]{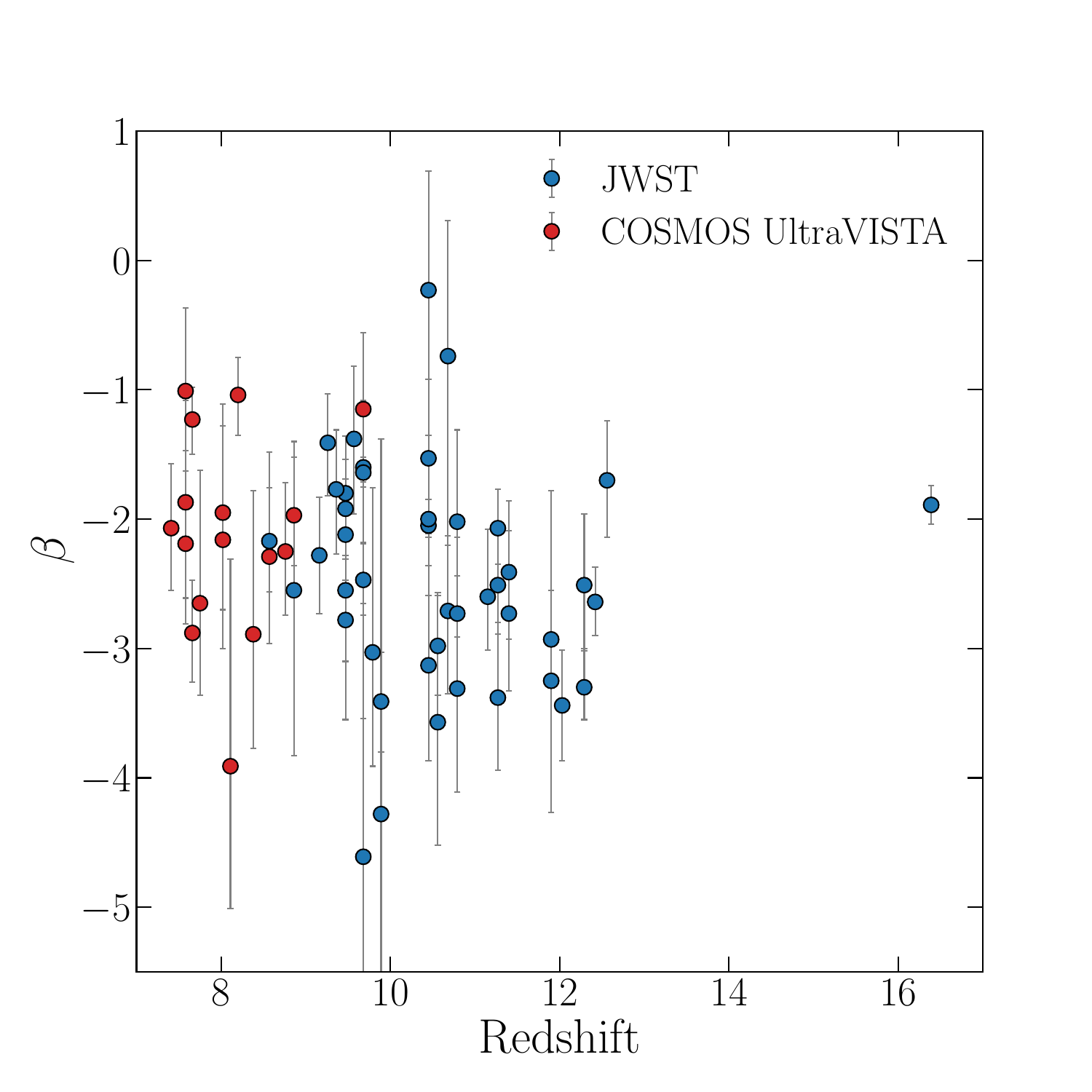}}
        \vspace{-0.25cm}
        \centerline{\includegraphics[width=\columnwidth]{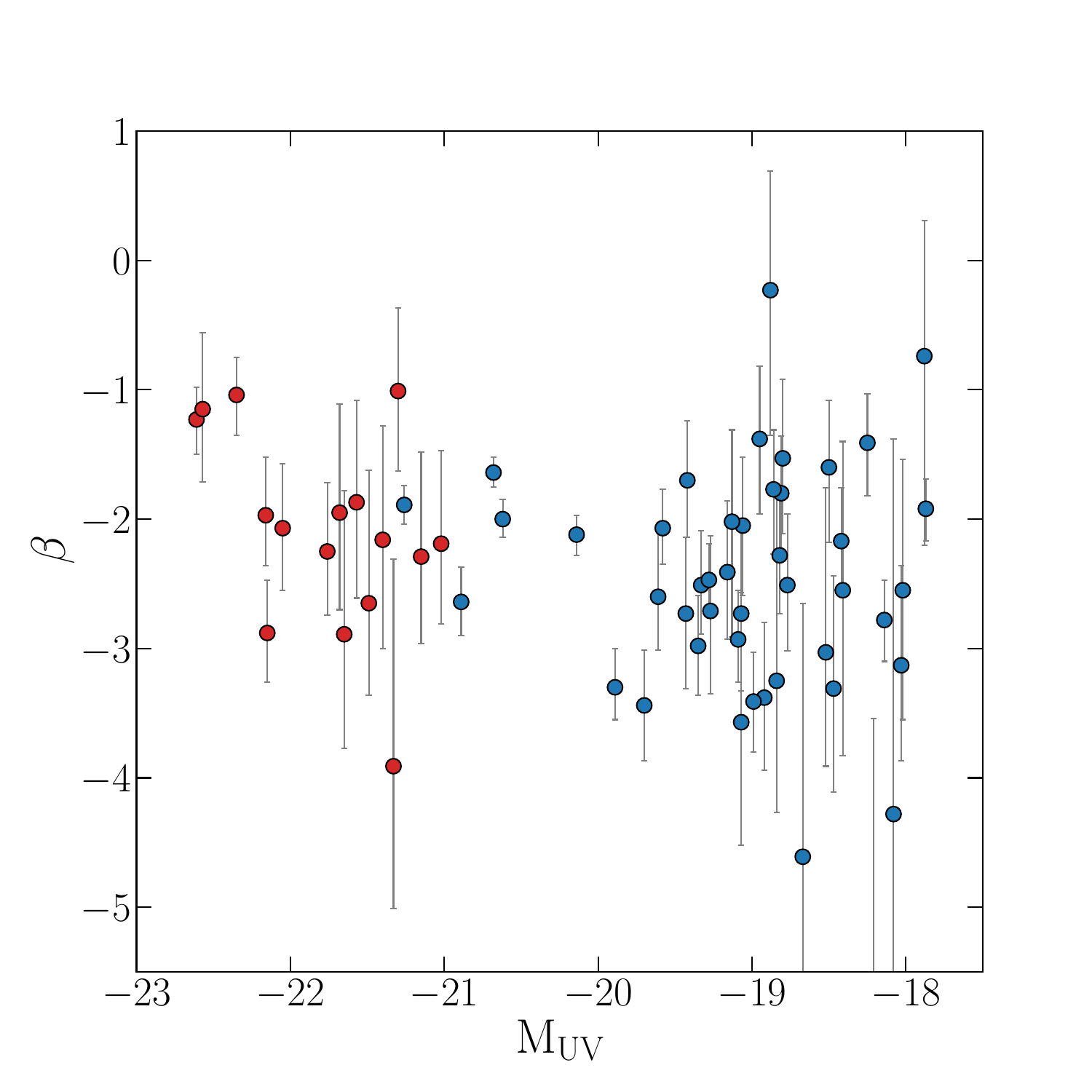}}
        \caption{Plots of UV continuum slope $\beta$ versus redshift (top) and versus absolute UV magnitude $M_{\rm UV}$ (bottom) for the galaxies in our JWST (blue) and COSMOS/UltraVISTA (red) sub-samples.}
        \label{fig:fig1}
    \end{figure}

In Fig.\,\ref{fig:fig1} we plot the $\beta$ values for our full JWST and COSMOS/UltraVISTA sample versus redshift, $z$, and absolute UV magnitude, $M_{\rm UV}$.  
As well as illustrating the typical $\beta$ values in our sample, the plots clearly demonstrate the power of combining JWST with ground-based surveys to probe a large dynamic range in both $z$ and, particularly, $M_{\rm UV}$.
The first point to note is the large scatter in observed $\beta$ values, which increases towards the faint luminosity limit in both samples.
This effect is seen most dramatically for the JWST sample, where values as extreme as $\beta < -4$ are recovered at $M_{\rm UV} \gtrsim - 19$.
However, the large error bars at these faint luminosities ($\sigma_{\beta} \simeq 1$ at $M_{\rm UV} > - 19$) suggests that this is predominantly a result of observational uncertainties.
Indeed, the preference for low-luminosity galaxies to be scattered blue is a well-known effect, caused by the fact that if a galaxy's flux is boosted into the detection band it will always be biased towards bluer UV slopes \citep{dunlop2013,rogers2013}. 
At the brightest UV luminosities in the JWST sample, where the constraints on individual $\beta$ estimates improve significantly, the scatter noticeably reduces and fewer ultra-blue ($\beta < -3)$ objects are seen.
We will discuss implications for the detection reliability of ultra-blue objects at faint luminosities in more detail in Sections \ref{subsec:ultra-blue-galaxies} and \ref{subsec:faint-end-bias}.

In the $\beta$ versus $z$ plot an increase in the scatter at $z\simeq9-11$ is apparent for the JWST sample. 
This is caused by a combination of (i) a larger number of intrinsically faint galaxies being detected in this redshift range, and (ii) a minimum in the number of filters covering rest-frame wavelengths $\lambda_{\rm rest} \leq 3000\, \text{\AA}$ (typically $\rm{N}_{\rm filt}=3$ at $z\lesssim10.5$ versus $\rm{N}_{\rm filt}=4$ at $z\gtrsim 10.5$, depending on the field).
Interestingly, one of our most robust $\beta$ estimates is the putative $z\simeq16.4$  galaxy candidate (CEERS 93316) reported in \cite{donnan2022}, which has $\beta=-1.9\pm0.15$.
This tight constraint is due, in part, to the excellent sampling of the rest-frame UV slope for this galaxy (it is covered by the F277W, F356W, F410M and F444W filters).
Promisingly, if these extremely high-redshift objects are confirmed - and if more are uncovered - JWST will be able to accurately constrain their UV continuum slopes thanks to the excellent coverage of the rest-frame UV continuum at $z>11$.

\begin{table}
    \centering
    \caption{Average $\beta$ values and standard errors derived for our full sample and in two bins of absolute UV magnitude. 
    The first column defines each sample in terms of $M_{\rm UV}$.
    In the second column we report the inverse-variance weighted mean and standard error of the individual $\beta$ values. 
    In the third column we report the median and $\sigma_{\rm MAD}$ of the individual $M_{\rm UV}$ values, where ${\sigma_{\rm MAD} = 1.483 \times \rm{MAD}}$ and MAD refers to the median absolute deviation.}
    \renewcommand{\arraystretch}{1.35}
    \begin{tabular}{l|c|c}
    \hline
    Sample & $\langle \beta \rangle$ & $\langle M_{\rm UV} \rangle$ \\
    \hline
      Full sample (all $M_{\rm UV}$)  & $-2.10 \pm 0.05$  &  $-19.3\pm1.3$ \\
      $M_{\rm UV} \leq -20.5$ & $-1.80\pm0.08$ & $-21.6\pm0.6$ \\
      $M_{\rm UV} >-20.5$ & $-2.32\pm0.07$ &  $-18.9\pm0.6$ \\
    \hline
    \end{tabular}
    \label{tab:beta}
\end{table}

It can be seen from Fig.\,\ref{fig:fig1} that the typical values of $\beta$ displayed by the galaxies in our $z~\simeq~8-16$ sample are somewhat bluer, but not obviously more extreme, than the typical values found at $z \leq 8$ with HST \citep[i.e., $\beta\simeq-2$;][]{dunlop2013}.
The distribution of points in Fig.\,\ref{fig:fig1} is consistent with being drawn from an underlying population with a relatively narrow \emph{intrinsic} distribution of $\beta$, with some evidence for a shallow trend towards bluer $\beta$ values at fainter $M_{\rm UV}$.
In Table \ref{tab:beta} we report the inverse-variance weighted mean $\beta$ value for our full sample, which we find to be $\langle \beta \rangle = -2.10 \pm 0.05$.
In this instance we preferred the weighted mean over the median so as not to be biased by the blue-scatter effect at faint luminosities (i.e., the blue-scattered galaxies are not down-weighted by their large uncertainties when taking the median).
Indeed, the median of the full sample is $\beta=-2.29 \pm 0.09$, where the uncertainty on the median is estimated using the median absolution deviation estimator ($\sigma_{\rm MAD} = 1.483 \times \rm{MAD}$).
As expected, the median estimate is bluer, although the formal difference is only at the $\simeq 2\sigma$ level.
Our sample average is in decent agreement with the median values reported at $z\simeq7-11$ in \cite{topping2022} ($\beta=-2.0$ at $z\simeq7$ and $\beta=-1.9$ at $z\simeq8-11$).

Adopting either the inverse-variance weighted mean or median, it is clear that our sample shows no evidence for significant evolution in the typical values of $\beta$ at $z>8$.
In fact, these early results imply that even the faintest galaxies that JWST has so far uncovered at $z \simeq 8-16$ have, \emph{on average}, UV colours no more extreme than the bluest galaxies in the local Universe (e.g., NGC 1705; $\beta=-2.46$, $M_{\rm UV} = -18$).

    \begin{figure}
     \vspace*{-0.1in}
        \centerline{\includegraphics[width=\columnwidth]{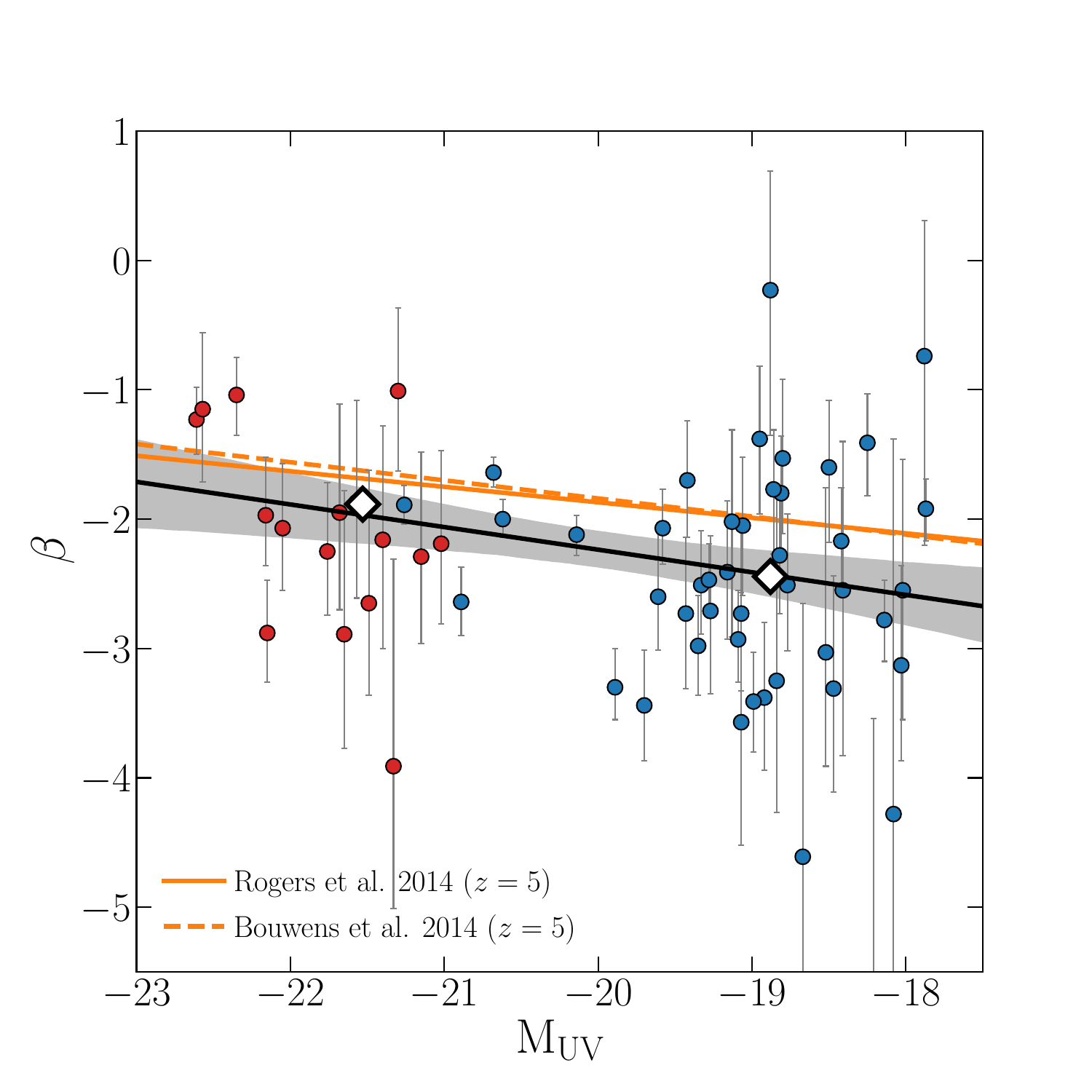}}
        \caption{A comparison between the $\beta$ versus $M_{\rm UV}$ relation at $z>8$ and previously-determined relations at lower redshift.
        The black solid line shows the best-fitting $\beta-M_{\rm UV}$ relation for our full sample which has a slope of $\rm{d}\beta/ \rm{d} M_{\rm UV} = -0.17 \pm 0.05$.
        The light-grey shaded region represents the $68$ per cent confidence interval around our best-fitting relation.
        The large diamond points are the inverse-variance weighted mean values of $\beta$ in the two bins of absolute UV magnitude given in Table \ref{tab:beta}.
        The orange dashed and dotted lines show the $z\simeq5$ relations from \citet{rogers2014} and \citet{bouwens2014} which have ${\rm{d}\beta / \rm{d} M_{\rm{UV}} = -0.12 \pm 0.02}$ and ${\rm{d}\beta / \rm{d} M_{\rm{UV}} = -0.14 \pm 0.02}$, respectively.
        }
        \label{fig:fig2}
    \end{figure}

\subsection{The $\mathbf{\beta-M_{\rm UV}}$ relation at $\mathbf{z>8}$}

The $\beta-M_{\rm UV}$ relation, often referred to as the colour-magnitude relation, encodes information on the dust and stellar population properties of galaxies as a function of their absolute UV magnitude.
A number of studies at $z\lesssim8$ have found strong evidence for a $\beta-M_{\rm UV}$ relation in which the UV continuum slopes of galaxies are bluer at fainter luminosities.
This has been used to argue that UV-faint galaxies are typically younger, less metal-enriched, and less dust-obscured than their brighter counterparts \citep[e.g.,][]{bouwens2014,rogers2014}.

Given the large dynamic range in $M_{\rm UV}$ provided by our combined JWST and COSMOS/UltraVISTA sample, we can examine early evidence for a $\beta-M_{\rm UV}$ relation  at ${z > 8}$.
In Table \ref{tab:beta}, we report the average $\beta$ values for our sample split into two magnitude bins divided at $M_{\rm UV}=-20.5$.
We find ${\langle \beta \rangle = -2.32 \pm 0.07}$ for the faint bin (median $M_{\rm UV}=-18.9$) and ${\langle \beta \rangle = -1.80 \pm 0.08}$ for the bright bin (median $M_{\rm UV}=-21.6$).
Our sample is therefore consistent with previous studies, with an evolution to redder colours in brighter galaxies.

Interestingly, we find that the formal best-fitting slope ${\beta-M_{\rm UV}}$ relation for our sample is fully consistent with relations derived at lower redshift (within $1\sigma$).
In Fig.\,\ref{fig:fig2} we plot the $\beta-M_{\rm UV}$ relations at $z\simeq5$ from \cite{rogers2014} and \cite{bouwens2014}, who both report a modest evolution in $\beta$ as a function of absolute UV magnitude, finding ${\rm{d}\beta/ \rm{d} M_{\rm UV} = -0.12 \pm 0.02}$ and ${\rm{d}\beta/ \rm{d} M_{\rm UV} = -0.14 \pm 0.02}$, respectively.
Fitting a similar colour-magnitude relation to our individual sources yields a best-fitting slope of $\rm{d}\beta/ \rm{d} M_{\rm UV} = -0.17 \pm 0.05$ (black solid line in Fig. \ref{fig:fig2})\footnote{Fitting to the two inverse-variance weighted mean values given in Table \ref{tab:beta} returns a formally steeper, but fully consistent, value of ${\rm{d}\beta/ \rm{d} M_{\rm UV} = -0.20 \pm 0.06}$.}.
The full best-fitting color-magnitude relation given by
\begin{equation}\label{eq:beta_muv}
    \beta = -0.17^{\,+0.05}_{-0.05} M_{\rm UV} - 5.40^{\,+1.18}_{-1.25}.
\end{equation}
Formally, $\chi^2/\nu=1.5$ for the best-fit model with respect to the data.
To obtain a rough estimate of the intrinsic scatter in the relation, we assumed the total variance in the data was a combination of the measurement error ($\sigma_{\rm m}$) and the intrinsic scatter ($\sigma_{\rm sc}$): ${\sigma_{\rm tot}^2=\sigma_{\rm m}^2+\sigma_{\rm sc}^2}$.
We found that the value of $\sigma_{\rm sc}$ that yielded $\chi^2/\nu \simeq 1$ was $\sigma_{\rm sc} \simeq 0.35$.
Interestingly, this value is again in good agreement with the result of \citet{rogers2014}, who estimated that the intrinsic scatter of the $\beta-M_{\rm UV}$ relation at $z \simeq 5$ increases from $\sigma_{\rm sc} \simeq 0.1$ at $M_{\rm UV}=-18$ to $\sigma_{\rm sc} \simeq 0.4$ at $M_{\rm UV}=-21$.

Our data also suggest that the normalisation of the relation has evolved such that, at higher redshifts, the typical $\beta$ values are bluer across the full $M_{\rm UV}$ range.
At bright magnitudes ($M_{\rm UV} \leq -20.5$) the inverse-variance weighted mean offset is $\langle \Delta \beta \rangle = -0.24 \pm 0.15$, increasing to $\langle \Delta \beta \rangle = -0.44 \pm 0.11$ at the faint end ($M_{\rm UV} > -20.5$).
The offset averaged across the full sample is $\langle \Delta \beta \rangle = -0.38 \pm 0.09$.

Evidence for a signal in these early datasets is encouraging, and future larger-area JWST surveys such as PRIMER (GO 1837) will clarify this situation in the near future.
These upcoming surveys will serve to both increase the sample size and fill the current magnitude gap at $-21 < M_{\rm UV} < -20$ where JWST can deliver excellent $\beta$ constraints.
Overall, the analysis of the $M_{\rm UV}-\beta$ relation further emphasises the main result of our analysis: although the galaxies at $z>8$ are generally bluer than their lower-redshift counterparts, \emph{on average}, the UV colours of our $z\simeq8-16$ galaxy sample are not dramatically bluer than bluest stellar populations observed at lower redshift, including sources at $z=0$.

\subsection{Evidence for ultra-blue objects ($\mathbf{\beta \simeq -3}$)?}\label{subsec:ultra-blue-galaxies}

Although the typical UV slopes in our sample appear to be no bluer than the bluest galaxies observed locally, ultra-blue objects (i.e., $\beta \leq -3$) may still exist within the population.
Indeed, \citet{topping2022} have recently identified two sources at $z \simeq 7$ with reportedly secure detections of $\beta \simeq -3$ from their investigation of the early CEERS NIRCam imaging data.
If confirmed, this would represent intriguing evidence for young, low metallicity stellar populations with ionizing continuum escape factions of $\simeq 100\%$ \citep[e.g.][]{robertson2010, chisholm2022}.

\begin{figure}
    \vspace*{-0.1in}
        \centerline{\includegraphics[width=\columnwidth]{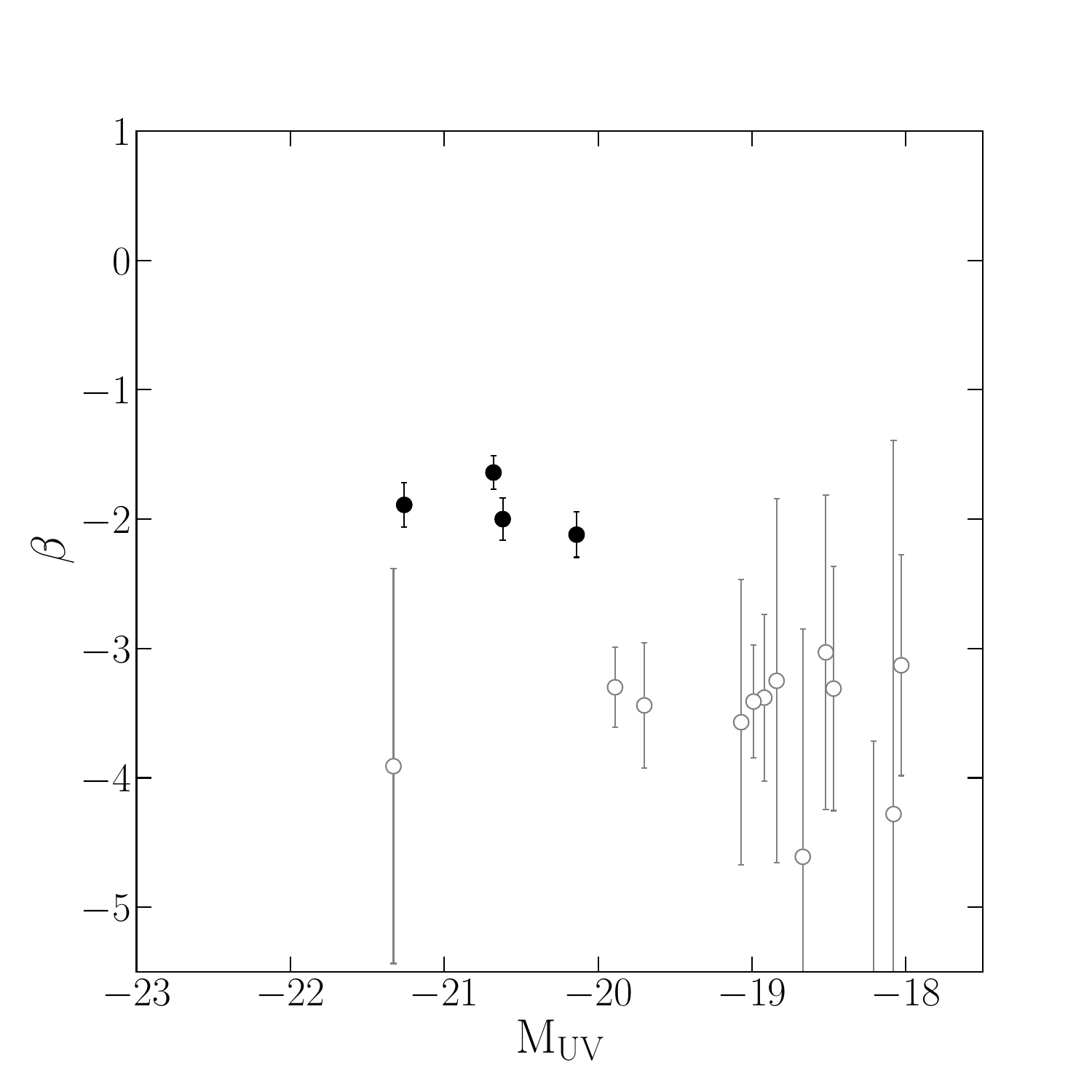}}
        \caption{Plot of $\beta$ versus $M_{\rm UV}$ for the objects with well constrained UV continuum slopes ($\sigma_{\beta} < 0.2$; black filled circles) and the objects with ${\beta \leq -3}$ (grey open circles). 
        We find no strong evidence for UV slopes as blue as $\beta \leq -3$ amongst those galaxies with robust measurements of $\beta$.
        All of the objects with formal $\beta \leq -3$ solutions are poorly constrained and consistent with the known blue bias in the $\beta$ scatter near the source-detection threshold.}
        \label{fig:fig3}
    \end{figure}

Our initial JWST and COSMOS/UltraVISTA sample does not provide convincing evidence for such objects. 
It can be seen in Fig.\,\ref{fig:fig3} that the majority of the galaxies in our sample with formal best fits of $\beta \leq -3$ have large uncertainties in the measurement of $\beta$.
In this case it is more likely that the galaxies have been scattered to blue values due to the known blue-bias in the $\beta$ scatter at faint luminosities \citep{dunlop2012, rogers2013}.
We provide a detailed discussion of this effect in Section \ref{subsec:faint-end-bias}.
In contrast, galaxies with well-constrained UV slopes, which we here define as those with an uncertainty of $\sigma_{\beta} \leq 0.2$\footnote{Although this definition is somewhat arbitrary, a galaxy with ${\beta \leq -3.0 \pm 0.2}$ would represent a $\gtrsim 5 \sigma$ deviation from the sample average of $\beta = -2.1$, and a $\simeq 3 \sigma$ deviation from the UV slope expected for dust free galaxy with a low escape fraction \citep[e.g., $\beta \simeq -2.4$;][]{cullen2017}, and would thus be a strong candidate for an exotic `ultra-blue' stellar population.} show no evidence for slopes bluer than $\beta \simeq -2.2$.
Overall, the results shown in Fig.\,\ref{fig:fig3} imply that the ultra-blue values we see in our sample are a result of statistical, rather than physical, effects.
    
The most plausible ultra-blue candidate in our sample is ID $32395\_2$ ($z=12.29$; $M_{\rm UV} = -19.89$), which has a formal best-fitting UV continuum slope of $\beta=-3.30^{\,+0.25}_{-0.30}$, a value consistent with the two ultra-blue candidates reported in \citet{topping2022}.
The UV slope for this object is formally a $4.8\sigma$ deviation from the sample average, but on closer inspection we find that this galaxy suffers from above-average systematic uncertainties.
For example, when fitting for both $\beta$ and $z$ (Section \ref{subsec:beta_fitting}) we find a large redshift offset from the \citet{donnan2022} estimate, with $\Delta z = -0.18$ (cf. the sample median of $\Delta z=-0.03$).
This shift in redshift results in a redder best-fitting UV continuum slope of $\beta=-3.1$.
Moreover, we find that ID $33593\_2$ is more sensitive to the chosen aperture size than the average galaxy in our sample.
Adopting smaller $0.36$ arcsec apertures yields $\beta=-2.99^{\,+0.25}_{-0.30}$.
This $0.36$ arcsec aperture value is consistent within $3.5\sigma$ of the sample average, and within $< 2\sigma$ of the UV continuum slope expected for a standard stellar populations with $Z \simeq 0.1 Z_{\odot}$ and $f_{\rm esc} = 0.0$ \citep[e.g., $\beta \simeq -2.4$;][]{cullen2017}.
Although this galaxy is almost certainly one of the bluest objects in our sample, the combined statistical and systematic uncertainties make it difficult to confirm it as a robust ultra-blue, $\beta < -3$, object.

Despite this, Fig.\,\ref{fig:fig3} does demonstrate that the currently-available JWST imaging can undoubtedly deliver well-constrained $\beta$ measurements ($\sigma_{\beta} \leq 0.2$) for galaxies as faint as $M_{\rm UV} \simeq -20$ at $z > 8$, and hence should be  able to uncover strong candidate $\beta \simeq -3$ objects at these absolute UV magnitudes should they exist \citep[e.g.,][]{topping2022}, despite the fact that no convincing candidates are found here in our current high-redshift galaxy sample. 
In a future study, exploiting data from upcoming, wider-area, JWST Cycle-1 imaging surveys, we intend to undertake a detailed analysis of the $\beta - M_{\rm UV}$ relation at $z \ge 7$ and attempt to robustly quantify the intrinsic scatter in the $\beta$ distribution.
    
\section{Discussion}\label{section:discussion}

We have presented the first estimate of the $\beta-M_{\rm UV}$ relation at $z>8$ using early JWST data.
We find that, on average, galaxies at these redshift are bluer than their lower redshift counterparts ($\Delta \langle \beta \rangle \simeq -0.4$ compared to $z=5$), but no bluer than the bluest objects uncovered at lower redshifts.
We do not find strong evidence for a significant population of ultra-blue $\beta < -3$ objects in our sample.

In this section we first provide a short discussion of our results, starting with an exploration of the well-know faint-end blue bias and its affect on our current sample.
We then compare our results to pre-JWST literature measurements at similar redshifts, as well as to predictions from a number of theoretical galaxy formation models.

\subsection{The blue $\mathrm{\beta}$ bias at faint magnitudes}\label{subsec:faint-end-bias}

A bias towards bluer values of $\beta$ at faint magnitudes is a well-known phenomenon that has been extensively documented in $HST$ studies \citep[e.g.][]{bouwens2010, dunlop2012, rogers2013}.
The bias occurs due the that fact that high-redshift galaxy candidates are typically selected using a photometric filter as close as possible to the Lyman break where the UV spectral energy distribution of young star-forming galaxies peaks.
At faint magnitudes, this favours the selection of objects whose photometry has been `up-scattered' in the short-wavelength detection band; these objects will naturally appear bluer than they actually are.

To investigate the magnitude of this effect in our JWST sample we ran a simple simulation.
We first constructed 20,000 simple power-law SEDs with a intrinsic UV slope of $\beta_{\rm int}=-2.1$ at the median redshift of the JWST galaxies ($z=10.5$).
The UV magnitudes were drawn uniformly within the range $-20.0 \leq M_{\rm UV} \leq -18.0$ and IGM attenuation was applied using the \citet{inoue2014} prescription.
Photometry was generated in each of the JWST filters and scattered according the the typical imaging depths (averaging the depths across multiple fields where appropriate).
The `observed' UV continuum slopes ($\beta_{\rm obs}$) were then recovered for the simulated galaxies using the same method applied to the real observations.

\begin{figure}
        \vspace*{-0.1in}
            \centerline{\includegraphics[width=\columnwidth]{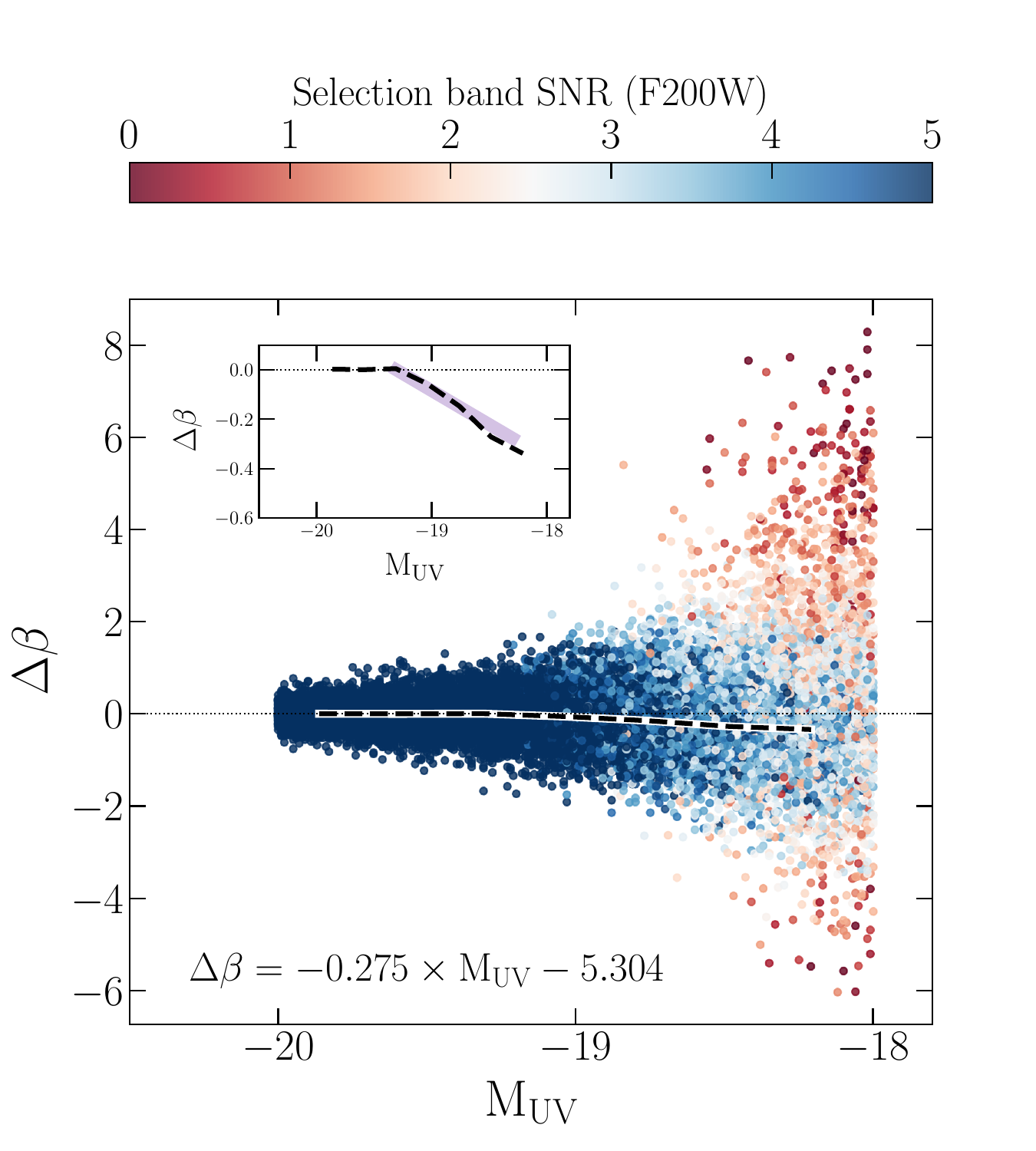}}
            \caption{Plot of the UV continuum slope bias ($\Delta \beta = \beta_{\rm obs} - \beta_{\rm int}$) as a function of $M_{\rm UV}$ for of 20,000 simulated galaxies at $z=10.5$ with $\beta_{\rm int}=-2.1$ (see Section \ref{subsec:faint-end-bias} for a description of the simulations).
            In the main panel, each data point represents an individual galaxy and is colour-coded according to the SNR in the F200W selection filter (the colour bar is saturated at $\rm{SNR}=5$ for clarity).
            The preference for galaxies with the highest SNR in the selection band to skew towards bluer observed $\beta$ (i.e., $\Delta \beta < 0$) is clearly visible at faint magnitudes (i.e., the colour asymmetry around the $\Delta \beta = 0$ line).
            The black dashed line shows the average bias for galaxies with SNR $\geq 5$ in F200W (i.e., mimicking our selection criteria) which tends to $\Delta \beta < 0$ at $M_{\rm UV} \gtrsim -19$.
            The inset panel shows a zoom-in of this average bias, with the purple line showing a linear fit to the relation.
            From our simple simulation, we find that the systematic bias in $\beta$ occurs for galaxies fainter than $M_{\rm UV}>-19.3$, and has the functional form ${\Delta \beta = -0.275 M_{\rm UV} - 5.304}$.}
            \label{fig:beta_sim}
    \end{figure}

The results of this simulation are shown in Fig. \ref{fig:beta_sim}.
Our results are consistent with the trends observed in previous works \citep[e.g,][]{rogers2013}.
We find that, at faint $M_{\rm UV}$, the galaxies with the highest signal-to-noise ratio in the detection band (F200W) are biased blue.
If we mimic the selection of our sample (i.e., requiring a $\geq 5 \sigma$ detection in F200W) we find that $\beta_{\rm int}$ is accurately recovered - on average - for galaxies brighter than $M_{\rm UV} \simeq -19.3$, but becomes increasingly biased to blue values at fainter magnitudes (black dashed line in Fig. \ref{fig:beta_sim}).
Fitting for this average bias (${\Delta \beta = \beta_{\rm obs} - \beta_{\rm int}}$) at  $M_{\rm UV} > -19.3$ we find that ${\Delta \beta = -0.275 M_{\rm UV} - 5.304}$.
Our results indicate that galaxies at the faint end of our sample (i.e., $M_{\rm UV} \simeq -18.5$) will have $\Delta \beta = -0.2$, on average.

Applying this average bias correction to the individual galaxies in our sample does not strongly affect our main results. 
We find that the recovered slope of the the ${\beta-M_{\rm UV}}$ relation becomes slightly shallower ($\rm{d}\beta/ \rm{d} M_{\rm UV} = -0.10 \pm 0.06$; still consistent with the \citet{rogers2014} and \citet{bouwens2014} slopes at $z=5$ within $1 \sigma$) and the inverse-variance weighed mean of the sample slightly redder ($\langle \beta \rangle = -2.05 \pm 0.05$), but both remain fully consistent with the original non-corrected values.
Nevertheless, as sample sizes increase, and the statistical uncertainties are reduced, this effect will clearly become more important, potentially requiring more sophisticated simulations including source injection/recovery, treatment of individual fields, and a consideration of aperture effects and redshift-dependent systematics.

\begin{figure*}
        \vspace*{-0.1in}
            \centerline{\includegraphics[width=\textwidth]{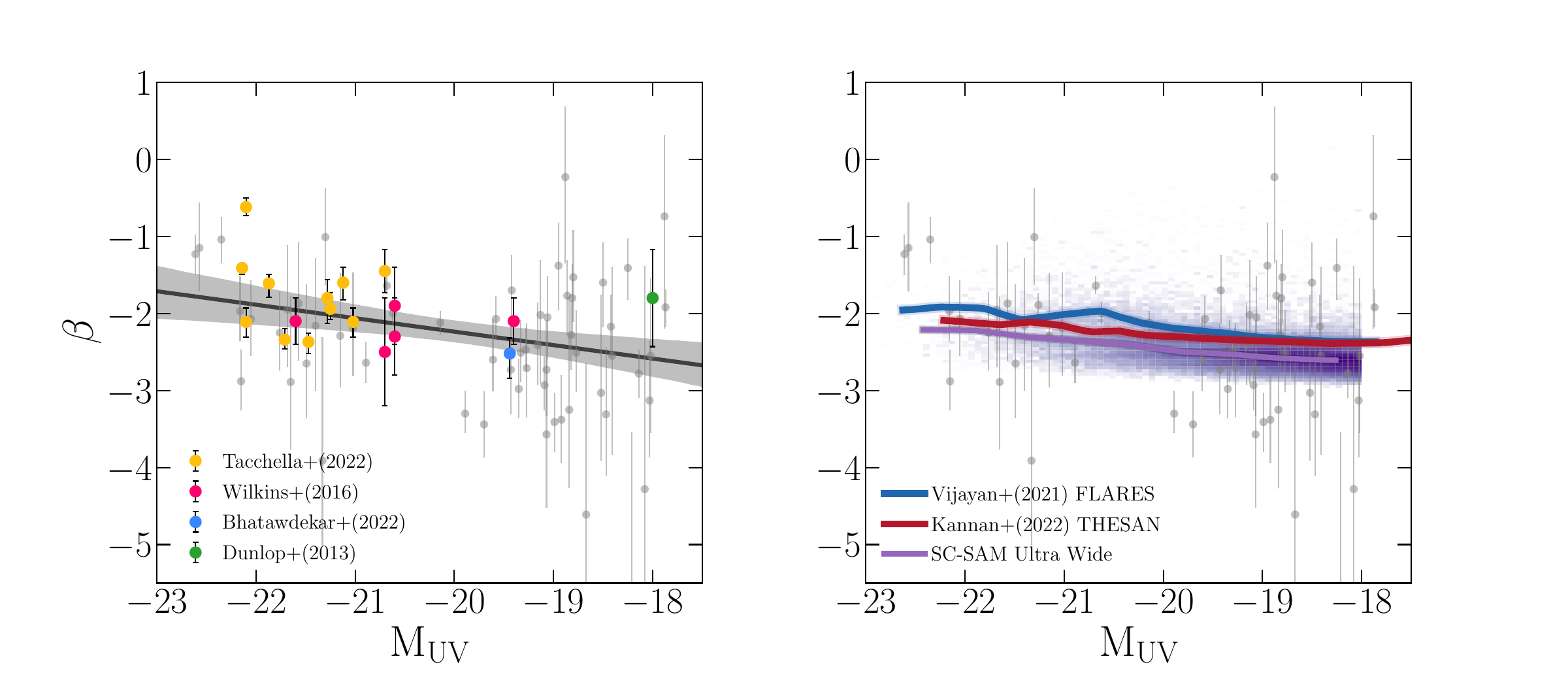}}
            \caption{A comparison of the $\beta-M_{\rm UV}$ relation at $z > 8$ with pre-JWST literature measurements and theoretical galaxy formation models. 
            In the left-hand panel we show a comparison of our individual data (light grey) and best-fitting relation (black line) to the four pre-JWST studies at $z\simeq9-10$ indicated in the legend \citep{dunlop2013,wilkins2016,bahtawdeker2021,tacchella2022}.
            In general, there is excellent agreement between our new measurements and the earlier literature data.
            This comparison also emphasises the power of these new JWST datasets, which have substantially increased the sample size at faint magnitude ($M_{\rm UV} \gtrsim -20.5$).
            In the right-hand panel we compare our data to three state-of-the-art simulations of galaxy formation: \textsc{sc-sam} \citep{yung2019a}, \textsc{flares} \citep{vijayan2021} and \textsc{thesan} \citep{kannan2022}.
            As discussed in Sec. \ref{subsec:thoretical_model_comparison}, the agreement is is qualitatively very good.
            The simulated $\beta-M_{\rm UV}$ relations have a similar normalization to our data, as well as similar predicted slopes (${\rm{d}\beta/ \rm{d} M_{\rm UV} \simeq 0.10-0.15}$).
            The purple 2D histogram shows the distribution of $\approx 150,000$ galaxies from \textsc{sc-sam} to illustrate the intrinsic scatter.
            In this simulation, the predicted intrinsic scatter varies from $\sigma_{\rm sc}=0.1$ at the faint end to $\sigma_{\rm sc}=0.3$ at the bright end, again in reasonable agreement with our estimate of $\sigma_{\rm sc}=0.35$.
            The good agreement with theoretical predictions suggests the galaxies in our sample are consistent with the young, metal-poor and moderately dust-reddened population predicted at $z>8$.}
            \label{fig:fig5}
    \end{figure*}

Fig. \ref{fig:fig5} also emphasises the fact that even in the magnitude regime where the average properties are accurately recovered, the scatter of any individual $\beta_{\rm obs}$ can still be substantial.
For example, at the magnitude of our most robust ultra-blue candidate ($M_{\rm UV} \simeq -20$; Section \ref{subsec:ultra-blue-galaxies}) the bias for individual objects in our simulation can be as large at $\Delta \beta = \pm 0.5$.
Ultimately, while promising ultra-blue candidates can be identified from data at these depths, robust confirmation will likely require either additional photometric tracers \cite[e.g., a lack of emission line signatures in the rest-frame optical photometry;][]{topping2022}, or deep spectroscopic follow-up.



\subsection{Comparison with pre-JWST results at $\mathbf{z \simeq 9-11}$}

Or results can be compared with a number of earlier works at ${z \simeq 9-11}$ undertaken prior to the launch of JWST.
These studies were based on either single-colour measurements with \textit{HST} (using the $J_{140}$ and $H_{160}$ filters), or a combination of \textit{HST} and \textit{Spitzer}/IRAC $3.6 \mu$m imaging.
Most samples consisted of UV-bright galaxies above the knee of the galaxy luminosity function \citep[i.e., $M_{\rm UV} \lesssim -20.5$;][]{bowler2020, donnan2022}.
At fainter magnitudes, galaxies were drawn from the ultra-deep \textit{HST} imaging in the \textit{Hubble Ultra Deep Field} \citep[UDF-12;][]{dunlop2013} or gravitational lensing fields.

Based on the UDF-12 dataset, \citet{dunlop2013} provided the first tentative estimate of $\langle \beta \rangle$ at $z=9$.
The \citet{dunlop2013} sample consisted of faint galaxies with $M_{\rm UV} \simeq -18$ and photometric measurements in the \textit{HST} $J_{140}$ and $H_{160}$ filters.
Using a single-colour estimate, \citet{dunlop2013} found $\langle \beta \rangle = -1.80 \pm 0.63$.
This value is larger (redder) than predicted by our best-fitting $\beta-M_{\rm UV}$ relation, but clearly consistent with our data within the substantial uncertainty (Fig. \ref{fig:fig5}).
Some degree of systematic offset is perhaps unsurprising given the limited rest-frame UV coverage of the \citet{dunlop2013} data.
Technically, the \citet{dunlop2013} measurement is an estimate of $\beta$ in the far-ultraviolet, whereas our measurement (and all others shown in Fig. \ref{fig:fig5})  span far- and near-ultraviolet wavelengths.
Indeed, if we restrict our data to similar rest-frame wavelengths ($\lesssim 1600${\AA}) we recover a inverse-variance weighed mean of ${\langle \beta \rangle = -1.90 \pm 0.10}$ for our JWST sample.
In that sense, the rest-frame FUV colors of the \citet{dunlop2013} sample are fully consistent with our new JWST sample.
However, the addition of longer-wavelength anchors at $\lambda > 2000${\AA} does favour a bluer ${\langle \beta \rangle}$.

\citet{wilkins2016} presented measurements of $\beta$ for five literature sources at $9.6 < z_{\rm phot} < 10.2$.
Their sample was comprised of four galaxies drawn from the GOODS-South and GOODS-North field \citep{oesch2014}, and one gravitatioanlly lensed source reported in \citet{zheng2012}.
The full sample spans the UV magnitude range $-21.6 \leq M_{\rm UV} \leq -19.4$.
UV continuum slopes for these sources were determined using a single colour measurement ($H_{\rm F160W}-3.6\mu \rm{m}$) probing the rest-frame UV continuum in the range $1200${\AA} $\lesssim \lambda \lesssim 3700${\AA}.
The five \citet{wilkins2016} sources are shown in the $\beta-M_{\rm UV}$ plane in the left-hand panel of Fig. \ref{fig:fig5} and are clearly in excellent agreement with our results.
Their sample average, estimated by stacking the photometry of the five individual sources, is $\langle \beta \rangle = -2.1 \pm 0.3$.
Restricting our sample to the same $M_{\rm UV}$ range as the \citet{wilkins2016} sample returns an inverse-variance weighted mean of $\langle \beta \rangle = -2.08 \pm 0.05$.

\citet{bahtawdeker2021} reported $\langle \beta \rangle = -2.52^{+0.32}_{-0.20}$ for eight lensed $z=9$ galaxies in the Frontier Field cluster ${\mathrm{MACS}\,\mathrm{J}0416.1-2403}$.
The median magnification-corrected absolute UV magnitude of their sample was $M_{\rm UV}=-19.44$. 
Stellar population models were fit to multi-wavelength photometry spanning $0.4-4.5\mu\mathrm{m}$ (i.e., rest-frame UV to optical wavelengths) and $\beta$ was calculated from the best-fitting model using the \citet{calzetti1994} windows.
The \citet{bahtawdeker2021} measurement is clearly in good agreement with our data; our best-fitting relation predicts $\langle \beta \rangle = -2.4$ at $M_{\rm UV}=-19.44$ (Fig. \ref{fig:fig5}).
Interestingly, \citet{bahtawdeker2021} also report a tentative color-magnitude slope of ${\rm{d}\beta/ \rm{d} M_{\rm UV} = -0.19 \pm 0.11}$ at $z=9$.
While the uncertainty on this value clearly significant, the formal best-fitting value is consistent our estimate.

Finally, \citet{tacchella2022} presented $\beta$ estimates for eleven bright ($M_{\rm UV} < -20.7$) galaxies at $z=9-11$, fitting to deep \textit{HST} and Spitzer/IRAC photometry and using a similar stellar population model-fitting technique to \citet{bahtawdeker2021}. 
Again, these measurements are fully consistent with our data, although in this case the comparison is primarily with our bright ground-based COSMOS/UltraVISTA sample (Fig. \ref{fig:fig5}).
The inverse variance weighed mean of the \cite{tacchella2022} sample is $\langle \beta \rangle = -1.63 \pm0.04$ compared to $\langle \beta \rangle = -1.79 \pm 0.12$ for our COSMOS sample.
The uncertainties on the \cite{tacchella2022} estimates are clearly smaller in this regime, primarily due to their deeper \textit{HST} photometry.
However, as discussed in detail by \cite{rogers2013}, it is worth bearing in mind that the uncertainties resulting from a template fitting approach can be artificially reduced by the limited parameter space afforded by population synthesis models (which act as a prior on the allowed values of $\beta$).

Overall, the comparison with previous work at $z\simeq9-11$ is clearly encouraging.
Fig. \ref{fig:fig5} also highlights the power the new JWST datasets, which are able to provide - for the first time - $\beta$ estimates for \emph{individual} galaxies at $M_{\rm UV} \gtrsim -20.5$ without the assistance of gravitational lensing.

\subsection{Comparison with galaxy-formation model predictions}\label{subsec:thoretical_model_comparison}

It is instructive to compare our results to the predictions of state-of-the-are galaxy formation models.
As our main comparison, we use the Santa Cruz semi-analytic model (\textsc{sc-sam}) for galaxy formation \citep{somerville2015, somerville2021}.
The \textsc{sc-sam} includes sophisticated prescriptions for cosmological accretion, gas cooling, star-formation, chemical enrichment and stellar and AGN feedback, and has been shown to successfully reproduce the global properties and scaling relations of the high-redshift galaxy population out to $z=10$ \citep[][]{yung2019a, yung2019b}.
The star-formation and chemical enrichment histories of the model galaxies are used to generate mock galaxy photometry based on the \citet{bruzual2003} stellar population synthesis models.
Dust attenuation is applied assuming a \citet{calzetti2000} attenuation law, with the rest-frame $V$-band dust attenuation calculated based on the surface density and metallicity of cold gas \citep{somerville2012}.
We refer interested readers to \citet{yung2022} for a detailed description of the model, including a flowchart illustrating the full internal workflow of the \textsc{sc-sam}. 

We obtained a sample of $\approx 150,000$ galaxies at ${8 < z < 10}$ with dust-attenuated absolute UV magnitudes in the range ${-23 < M_{\rm UV} < -18}$ from the \textsc{sc-sam} ultra-wide lightcone \citep[covering 2 deg$^2$;][]{yung2022b}\footnote{http://flathub.flatironinstitute.org/group/sam-forecasts}.
These galaxies have predicted stellar masses in the range $10^7 - 10^{10} M_{\odot}$ and stellar metallicities in the range $0.01-1 Z_{\odot}$.
The median mass and metallicity of the simulated sample is $M_{\star} \simeq 10^{8.5}M_{\odot}$ and $Z_{\star} \simeq 0.1Z_{\odot}$.
The mean stellar age of the sample is $\simeq 100$ Myr.
We calculated an observed $\beta$ for each galaxy (i.e., after the application of dust reddening) by fitting a power-law to the noise-free mock photometry.
The predicted $\beta-M_{\rm UV}$ relation (observed) is shown in the right-hand panel of Fig. \ref{fig:fig5}, where it can be seen that the overall normalization of the relation is in good agreement with our data.
The slope of the \textsc{sc-sam} relation (${\rm{d}\beta/ \rm{d} M_{\rm UV} \simeq -0.15}$) is also fully consistent with our estimate.
The 2D histogram in Fig. \ref{fig:fig5} shows the intrinsic scatter, which increases from $\sigma_{\rm sc}=0.1$ at $M_{\rm UV}=-18$ to $\sigma_{\rm sc}=0.3$ at $M_{\rm UV}=-22$, again in reasonable agreement with our estimate of $\sigma_{\rm sc}=0.35$.

Both the shape and scatter of the $\beta-M_{\rm UV}$ relation in the \textsc{sc-sam} are driven by UV dust attenuation ($A_{\rm UV}$).
This is unsurprising, as the intrinsic low-order shape of the UV continuum is not strongly metallicity-dependent \citep[][]{cullen2019}, and at high redshifts the of effect stellar population age is limited by the young age of the Universe \citep{tacchella2022}.
Across the full range in $M_{\rm UV}$, the median UV attenuation increases from $A_{\rm UV}=0.05$ to $A_{\rm UV}=0.64$ (i.e.,  galaxies at the bright end in the \textsc{sc-sam} suffer a factor $\simeq$ $2$ decrease in their intrinsic UV flux).
The increase in scatter at the bright end is also driven by dust, with a larger range of $A_{\rm UV}$ at bright magnitudes.
At ${M_{\rm UV}=-22}$, the standard deviation of the $A_{\rm UV}$ distribution is ${\sigma_{A_{\rm UV}}=0.4}$, compared to ${\sigma_{A_{\rm UV}}=0.1}$ at $M_{\rm UV}=-18$.

In the right-hand panel of Fig. \ref{fig:fig5} we also show $\beta-M_{\rm UV}$ predictions at $z=9$ from the \textsc{flares} \citep{lovell2021,vijayan2021} and \textsc{thesan} \citep{kannan2022,smith2022} hydrodymanical simulations.
In both simulations the ISM metallicity is combined with an assumed dust-to-metal ratio to determine absolute dust attenuation.
To compute the wavelength-dependent attenuation, \textsc{flares} employ a \citet{charlotfall2000} dust model with an attenuation curve somewhat stepper than the \citet{calzetti2000} law, whereas \textsc{thesan} implement full dust radiative transfer using \textsc{skirt} \citep{camps2020}.
Both simulations predict slightly larger $\beta$ values compared to the \textsc{sc-sam} model and slightly shallower $\beta-M_{\rm UV}$ slopes (${\rm{d}\beta/ \rm{d} M_{\rm UV} \simeq -0.1}$).
However, neither are obviously incompatible with our data and paint the same basic picture of a uniformly blue ($\beta \lesssim -2$) galaxy population at $z>8$.
Overall, the reasonable agreement between the different simulations, and their consistency with early JWST observations, is encouraging.

Based on this comparison, we can infer that our early $z>8$ data are qualitatively consistent with the young, low-metallicity, moderately dust-reddened stellar populations predicted by theoretical models.



\section{Conclusions}\label{section:conclusion}

We have measured the rest-frame ultraviolet (UV) continuum slopes ($\beta$) of 61 galaxies in the redshift range $8 < z < 16$, using a combination of {\it James Webb Space Telescope} (JWST) ERO and ERS NIRcam imaging and ground-based near-infrared imaging of the COSMOS/UltraVISTA field. The primary aim of this analysis is to determine whether there is any evidence for an evolution in the typical UV colours of the new population of $z>8$ galaxies being uncovered by JWST.
We present early estimates of the \emph{average} values of $\beta$ at these redshifts and, using the large dynamic range in UV luminosity enabled by the combination of JWST and ground-based imaging, we investigate evidence for a $\beta-M_{\rm UV}$ relation in our sample.
Our main conclusions can be summarized as follows.

\begin{enumerate}

    \item Using a power-law fitting technique, we find that our full sample displays a weighted mean value of $\langle \beta \rangle = -2.10 \pm 0.05$, with a corresponding median value of $\beta=-2.29\pm 0.09$.
    This result implies that  even the faintest galaxies that JWST has so far uncovered at $z \simeq 8-16$ have, on average, UV colours no more extreme than the bluest galaxies in the local Universe (e.g., NGC 1705; $\beta = -2.46$).

    \item We find evidence for a $\beta-M_{\rm UV}$ relation in our sample, such that brighter UV galaxies display somewhat redder UV slopes (${\rm{d}\beta/ \rm{d} M_{\rm UV} = -0.17 \pm 0.06}$).
    This value is consistent with the slope derived at $z\simeq5$ by \cite{rogers2014} and \cite{bouwens2014} (${\rm{d}\beta/ \rm{d} M_{\rm UV} = -0.12 \pm 0.02}$ and ${\rm{d}\beta/ \rm{d} M_{\rm UV} = -0.14 \pm 0.02}$, respectively).
    Compared to the the $z\simeq5$ relations, we find that the galaxies in our sample are bluer than their $z\simeq 5$ counterparts.
    The inverse-variance weighted mean offset for the full sample is $\langle \Delta \beta \rangle = -0.38 \pm 0.09$.

    \item Examining the $\beta$ estimates for individual galaxies in our sample, we find no strong evidence for galaxies with ultra-blue UV slopes that would indicate extreme stellar populations (i.e., $\beta \leq -3$).
    The majority of galaxies in our sample with $\beta \leq -3$ have significant uncertainties, and appear to be consistent with the well-known blue bias in the $\beta$ scatter for sources near to the source detection threshold.

    
\end{enumerate}

Overall, we find that the new population of $8 < z < 16$ galaxies being uncovered with JWST is on average bluer than the population at lower redshfits, but does not show evidence of a substantial fraction of ultra-blue $\beta < -3$ sources.
Indeed, our sample is consistent with the young ($\simeq 100$ Myr), metal-poor ($\simeq 0.1 Z_{\odot}$) and moderately dust-reddened galaxies predicted by current theoretical galaxy formation models at these redshifts.

\section*{Acknowledgements}

F. Cullen and T. M. Stanton acknowledge support from a UKRI Frontier Research Guarantee Grant (PI Cullen; grant reference EP/X021025/1).
R.\,J. McLure, D.\,J. McLeod, J.\,S.~Dunlop, C. Donnan, R. Begley and M.\,L. Hamadouche, acknowledge the support of the Science and Technology Facilities Council. 
A.\,C. Carnall thanks the Leverhulme Trust for their support via a Leverhulme Early Career Fellowship. 
R.\,A.\,A. Bowler acknowledges support from an STFC Ernest Rutherford Fellowship (grant number ST/T003596/1).

We would like to thank Aaron Yung for kindly helping with the \textsc{sc-sam} lightcone data. We would also like to thank Dan Magee for useful discussions and support in the data reduction process.
This work is based on observations collected at the European Southern Observatory under ESO programme ID 179.A-2005 and 198.A-2003 and on data products produced by CALET and the Cambridge Astronomy Survey Unit on behalf of the UltraVISTA consortium.
For the purpose of open access, the author has applied a Creative Commons Attribution (CC BY) licence to any Author Accepted Manuscript version arising from this submission.

\vspace*{-0.15in}

\section*{Data Availability}

All JWST and HST data products are available via the Mikulski Archive for Space Telescopes (\url{https://mast.stsci.edu}). UltraVISTA DR5 will shortly be made available through ESO. Additional data products are available from the authors upon reasonable request.



\vspace*{-0.15in}

\bibliographystyle{mnras}
\bibliography{jwst_beta} 

\begin{thebibliography}{}
\makeatletter
\relax
\def\mn@urlcharsother{\let\do\@makeother \do\$\do\&\do\#\do\^\do\_\do\%\do\~}
\def\mn@doi{\begingroup\mn@urlcharsother \@ifnextchar [ {\mn@doi@}
  {\mn@doi@[]}}
\def\mn@doi@[#1]#2{\def\@tempa{#1}\ifx\@tempa\@empty \href
  {http://dx.doi.org/#2} {doi:#2}\else \href {http://dx.doi.org/#2} {#1}\fi
  \endgroup}
\def\mn@eprint#1#2{\mn@eprint@#1:#2::\@nil}
\def\mn@eprint@arXiv#1{\href {http://arxiv.org/abs/#1} {{\tt arXiv:#1}}}
\def\mn@eprint@dblp#1{\href {http://dblp.uni-trier.de/rec/bibtex/#1.xml}
  {dblp:#1}}
\def\mn@eprint@#1:#2:#3:#4\@nil{\def\@tempa {#1}\def\@tempb {#2}\def\@tempc
  {#3}\ifx \@tempc \@empty \let \@tempc \@tempb \let \@tempb \@tempa \fi \ifx
  \@tempb \@empty \def\@tempb {arXiv}\fi \@ifundefined
  {mn@eprint@\@tempb}{\@tempb:\@tempc}{\expandafter \expandafter \csname
  mn@eprint@\@tempb\endcsname \expandafter{\@tempc}}}

\bibitem[\protect\citeauthoryear{{Adams} et~al.,}{{Adams}
  et~al.}{2023}]{adams2022}
{Adams} N.~J.,  et~al., 2023, \mn@doi [\mnras] {10.1093/mnras/stac3347}, \href
  {https://ui.adsabs.harvard.edu/abs/2023MNRAS.518.4755A} {518, 4755}

\bibitem[\protect\citeauthoryear{{Aihara} et~al.,}{{Aihara}
  et~al.}{2019}]{aihara2019}
{Aihara} H.,  et~al., 2019, \mn@doi [\pasj] {10.1093/pasj/psz103}, \href
  {https://ui.adsabs.harvard.edu/abs/2019PASJ...71..114A} {71, 114}

\bibitem[\protect\citeauthoryear{{Atek}, {Shuntov}, {Furtak}, {Richard},
  {Kneib}, {Mahler Adi Zitrin}  \& {McCracken Clotilde Laigle St{\'e}phane
  Charlot}}{{Atek} et~al.}{2022}]{atek2022}
{Atek} H.,  {Shuntov} M.,  {Furtak} L.~J.,  {Richard} J.,  {Kneib} J.-P.,
  {Mahler Adi Zitrin} G.,   {McCracken Clotilde Laigle St{\'e}phane Charlot}
  H.~J.,  2022, arXiv e-prints, \href
  {https://ui.adsabs.harvard.edu/abs/2022arXiv220712338A} {p. arXiv:2207.12338}

\bibitem[\protect\citeauthoryear{{Bertin} \& {Arnouts}}{{Bertin} \&
  {Arnouts}}{1996}]{Bertin1996}
{Bertin} E.,  {Arnouts} S.,  1996, \mn@doi [\aaps] {10.1051/aas:1996164}, \href
  {https://ui.adsabs.harvard.edu/abs/1996A&AS..117..393B} {117, 393}

\bibitem[\protect\citeauthoryear{{Bhatawdekar} \& {Conselice}}{{Bhatawdekar} \&
  {Conselice}}{2021}]{bahtawdeker2021}
{Bhatawdekar} R.,  {Conselice} C.~J.,  2021, \mn@doi [\apj]
  {10.3847/1538-4357/abdd3f}, \href
  {https://ui.adsabs.harvard.edu/abs/2021ApJ...909..144B} {909, 144}

\bibitem[\protect\citeauthoryear{{Bouwens} et~al.,}{{Bouwens}
  et~al.}{2010}]{bouwens2010}
{Bouwens} R.~J.,  et~al., 2010, \mn@doi [\apjl] {10.1088/2041-8205/708/2/L69},
  \href {https://ui.adsabs.harvard.edu/abs/2010ApJ...708L..69B} {708, L69}

\bibitem[\protect\citeauthoryear{{Bouwens} et~al.,}{{Bouwens}
  et~al.}{2014}]{bouwens2014}
{Bouwens} R.~J.,  et~al., 2014, \mn@doi [\apj] {10.1088/0004-637X/793/2/115},
  \href {https://ui.adsabs.harvard.edu/abs/2014ApJ...793..115B} {793, 115}

\bibitem[\protect\citeauthoryear{{Bowler}, {Jarvis}, {Dunlop}, {McLure},
  {McLeod}, {Adams}, {Milvang-Jensen}  \& {McCracken}}{{Bowler}
  et~al.}{2020}]{bowler2020}
{Bowler} R.~A.~A.,  {Jarvis} M.~J.,  {Dunlop} J.~S.,  {McLure} R.~J.,  {McLeod}
  D.~J.,  {Adams} N.~J.,  {Milvang-Jensen} B.,   {McCracken} H.~J.,  2020,
  \mn@doi [\mnras] {10.1093/mnras/staa313}, \href
  {https://ui.adsabs.harvard.edu/abs/2020MNRAS.493.2059B} {493, 2059}

\bibitem[\protect\citeauthoryear{{Brammer}, {van Dokkum}  \& {Coppi}}{{Brammer}
  et~al.}{2008}]{brammer2008}
{Brammer} G.~B.,  {van Dokkum} P.~G.,   {Coppi} P.,  2008, \mn@doi [\apj]
  {10.1086/591786}, \href
  {https://ui.adsabs.harvard.edu/abs/2008ApJ...686.1503B} {686, 1503}

\bibitem[\protect\citeauthoryear{{Bruzual} \& {Charlot}}{{Bruzual} \&
  {Charlot}}{2003}]{bruzual2003}
{Bruzual} G.,  {Charlot} S.,  2003, \mn@doi [\mnras]
  {10.1046/j.1365-8711.2003.06897.x}, \href
  {https://ui.adsabs.harvard.edu/abs/2003MNRAS.344.1000B} {344, 1000}

\bibitem[\protect\citeauthoryear{{Calzetti}, {Kinney}  \&
  {Storchi-Bergmann}}{{Calzetti} et~al.}{1994}]{calzetti1994}
{Calzetti} D.,  {Kinney} A.~L.,   {Storchi-Bergmann} T.,  1994, \mn@doi [\apj]
  {10.1086/174346}, \href
  {https://ui.adsabs.harvard.edu/abs/1994ApJ...429..582C} {429, 582}

\bibitem[\protect\citeauthoryear{{Calzetti}, {Armus}, {Bohlin}, {Kinney},
  {Koornneef}  \& {Storchi-Bergmann}}{{Calzetti} et~al.}{2000}]{calzetti2000}
{Calzetti} D.,  {Armus} L.,  {Bohlin} R.~C.,  {Kinney} A.~L.,  {Koornneef} J.,
   {Storchi-Bergmann} T.,  2000, \mn@doi [\apj] {10.1086/308692}, \href
  {https://ui.adsabs.harvard.edu/abs/2000ApJ...533..682C} {533, 682}

\bibitem[\protect\citeauthoryear{{Camps} \& {Baes}}{{Camps} \&
  {Baes}}{2020}]{camps2020}
{Camps} P.,  {Baes} M.,  2020, \mn@doi [Astronomy and Computing]
  {10.1016/j.ascom.2020.100381}, \href
  {https://ui.adsabs.harvard.edu/abs/2020A&C....3100381C} {31, 100381}

\bibitem[\protect\citeauthoryear{{Castellano} et~al.,}{{Castellano}
  et~al.}{2022}]{castellano2022}
{Castellano} M.,  et~al., 2022, arXiv e-prints, \href
  {https://ui.adsabs.harvard.edu/abs/2022arXiv220709436C} {p. arXiv:2207.09436}

\bibitem[\protect\citeauthoryear{{Charlot} \& {Fall}}{{Charlot} \&
  {Fall}}{2000}]{charlotfall2000}
{Charlot} S.,  {Fall} S.~M.,  2000, \mn@doi [\apj] {10.1086/309250}, \href
  {https://ui.adsabs.harvard.edu/abs/2000ApJ...539..718C} {539, 718}

\bibitem[\protect\citeauthoryear{{Chisholm} et~al.,}{{Chisholm}
  et~al.}{2022}]{chisholm2022}
{Chisholm} J.,  et~al., 2022, \mn@doi [\mnras] {10.1093/mnras/stac2874}, \href
  {https://ui.adsabs.harvard.edu/abs/2022MNRAS.517.5104C} {517, 5104}

\bibitem[\protect\citeauthoryear{{Cullen}, {McLure}, {Khochfar}, {Dunlop}  \&
  {Dalla Vecchia}}{{Cullen} et~al.}{2017}]{cullen2017}
{Cullen} F.,  {McLure} R.~J.,  {Khochfar} S.,  {Dunlop} J.~S.,   {Dalla
  Vecchia} C.,  2017, \mn@doi [\mnras] {10.1093/mnras/stx1451}, \href
  {https://ui.adsabs.harvard.edu/abs/2017MNRAS.470.3006C} {470, 3006}

\bibitem[\protect\citeauthoryear{{Cullen} et~al.,}{{Cullen}
  et~al.}{2019}]{cullen2019}
{Cullen} F.,  et~al., 2019, \mn@doi [\mnras] {10.1093/mnras/stz1402}, \href
  {https://ui.adsabs.harvard.edu/abs/2019MNRAS.487.2038C} {487, 2038}

\bibitem[\protect\citeauthoryear{{Curtis-Lake} et~al.,}{{Curtis-Lake}
  et~al.}{2022}]{curtis-lake2022}
{Curtis-Lake} E.,  et~al., 2022, arXiv e-prints, \href
  {https://ui.adsabs.harvard.edu/abs/2022arXiv221204568C} {p. arXiv:2212.04568}

\bibitem[\protect\citeauthoryear{{Donnan} et~al.,}{{Donnan}
  et~al.}{2022}]{donnan2022}
{Donnan} C.~T.,  et~al., 2022, \mn@doi [\mnras] {10.1093/mnras/stac3472}, \href
  {https://ui.adsabs.harvard.edu/abs/2022MNRAS.tmp.3239D} {}

\bibitem[\protect\citeauthoryear{{Dunlop}, {McLure}, {Robertson}, {Ellis},
  {Stark}, {Cirasuolo}  \& {de Ravel}}{{Dunlop} et~al.}{2012}]{dunlop2012}
{Dunlop} J.~S.,  {McLure} R.~J.,  {Robertson} B.~E.,  {Ellis} R.~S.,  {Stark}
  D.~P.,  {Cirasuolo} M.,   {de Ravel} L.,  2012, \mn@doi [\mnras]
  {10.1111/j.1365-2966.2011.20102.x}, \href
  {https://ui.adsabs.harvard.edu/abs/2012MNRAS.420..901D} {420, 901}

\bibitem[\protect\citeauthoryear{{Dunlop} et~al.,}{{Dunlop}
  et~al.}{2013}]{dunlop2013}
{Dunlop} J.~S.,  et~al., 2013, \mn@doi [\mnras] {10.1093/mnras/stt702}, \href
  {https://ui.adsabs.harvard.edu/abs/2013MNRAS.432.3520D} {432, 3520}

\bibitem[\protect\citeauthoryear{{Euclid Collaboration} et~al.,}{{Euclid
  Collaboration} et~al.}{2022}]{moneti2022}
{Euclid Collaboration} et~al., 2022, \mn@doi [\aap]
  {10.1051/0004-6361/202142361}, \href
  {https://ui.adsabs.harvard.edu/abs/2022A&A...658A.126E} {658, A126}

\bibitem[\protect\citeauthoryear{{Finkelstein} et~al.,}{{Finkelstein}
  et~al.}{2012}]{finkelstein2012}
{Finkelstein} S.~L.,  et~al., 2012, \mn@doi [\apj]
  {10.1088/0004-637X/756/2/164}, \href
  {https://ui.adsabs.harvard.edu/abs/2012ApJ...756..164F} {756, 164}

\bibitem[\protect\citeauthoryear{{Finkelstein} et~al.,}{{Finkelstein}
  et~al.}{2022}]{finkelstein2022}
{Finkelstein} S.~L.,  et~al., 2022, arXiv e-prints, \href
  {https://ui.adsabs.harvard.edu/abs/2022arXiv220712474F} {p. arXiv:2207.12474}

\bibitem[\protect\citeauthoryear{{Harikane} et~al.,}{{Harikane}
  et~al.}{2022}]{harikane2022}
{Harikane} Y.,  et~al., 2022, arXiv e-prints, \href
  {https://ui.adsabs.harvard.edu/abs/2022arXiv220801612H} {p. arXiv:2208.01612}

\bibitem[\protect\citeauthoryear{{Hudelot} et~al.,}{{Hudelot}
  et~al.}{2012}]{hudelot2012}
{Hudelot} P.,  et~al., 2012, VizieR Online Data Catalog, \href
  {https://ui.adsabs.harvard.edu/abs/2012yCat.2317....0H} {p. II/317}

\bibitem[\protect\citeauthoryear{{Inoue}, {Shimizu}, {Iwata}  \&
  {Tanaka}}{{Inoue} et~al.}{2014}]{inoue2014}
{Inoue} A.~K.,  {Shimizu} I.,  {Iwata} I.,   {Tanaka} M.,  2014, \mn@doi
  [\mnras] {10.1093/mnras/stu936}, \href
  {https://ui.adsabs.harvard.edu/abs/2014MNRAS.442.1805I} {442, 1805}

\bibitem[\protect\citeauthoryear{{Kannan}, {Smith}, {Garaldi}, {Shen},
  {Vogelsberger}, {Pakmor}, {Springel}  \& {Hernquist}}{{Kannan}
  et~al.}{2022}]{kannan2022}
{Kannan} R.,  {Smith} A.,  {Garaldi} E.,  {Shen} X.,  {Vogelsberger} M.,
  {Pakmor} R.,  {Springel} V.,   {Hernquist} L.,  2022, \mn@doi [\mnras]
  {10.1093/mnras/stac1557}, \href
  {https://ui.adsabs.harvard.edu/abs/2022MNRAS.514.3857K} {514, 3857}

\bibitem[\protect\citeauthoryear{{Lovell}, {Vijayan}, {Thomas}, {Wilkins},
  {Barnes}, {Irodotou}  \& {Roper}}{{Lovell} et~al.}{2021}]{lovell2021}
{Lovell} C.~C.,  {Vijayan} A.~P.,  {Thomas} P.~A.,  {Wilkins} S.~M.,  {Barnes}
  D.~J.,  {Irodotou} D.,   {Roper} W.,  2021, \mn@doi [\mnras]
  {10.1093/mnras/staa3360}, \href
  {https://ui.adsabs.harvard.edu/abs/2021MNRAS.500.2127L} {500, 2127}

\bibitem[\protect\citeauthoryear{{McCracken} et~al.,}{{McCracken}
  et~al.}{2012}]{mccracken2012}
{McCracken} H.~J.,  et~al., 2012, \mn@doi [\aap] {10.1051/0004-6361/201219507},
  \href {https://ui.adsabs.harvard.edu/abs/2012A&A...544A.156M} {544, A156}

\bibitem[\protect\citeauthoryear{{McLure} et~al.,}{{McLure}
  et~al.}{2011}]{mclure2011}
{McLure} R.~J.,  et~al., 2011, \mn@doi [\mnras]
  {10.1111/j.1365-2966.2011.19626.x}, \href
  {https://ui.adsabs.harvard.edu/abs/2011MNRAS.418.2074M} {418, 2074}

\bibitem[\protect\citeauthoryear{{McLure} et~al.,}{{McLure}
  et~al.}{2018}]{mclure2018}
{McLure} R.~J.,  et~al., 2018, \mn@doi [\mnras] {10.1093/mnras/sty522}, \href
  {https://ui.adsabs.harvard.edu/abs/2018MNRAS.476.3991M} {476, 3991}

\bibitem[\protect\citeauthoryear{{Naidu} et~al.,}{{Naidu}
  et~al.}{2022}]{naidu2022}
{Naidu} R.~P.,  et~al., 2022, arXiv e-prints, \href
  {https://ui.adsabs.harvard.edu/abs/2022arXiv220709434N} {p. arXiv:2207.09434}

\bibitem[\protect\citeauthoryear{{Oesch} et~al.,}{{Oesch}
  et~al.}{2014}]{oesch2014}
{Oesch} P.~A.,  et~al., 2014, \mn@doi [\apj] {10.1088/0004-637X/786/2/108},
  \href {https://ui.adsabs.harvard.edu/abs/2014ApJ...786..108O} {786, 108}

\bibitem[\protect\citeauthoryear{{Oke}}{{Oke}}{1974}]{oke1974}
{Oke} J.~B.,  1974, \mn@doi [\apjs] {10.1086/190287}, \href
  {https://ui.adsabs.harvard.edu/abs/1974ApJS...27...21O} {27, 21}

\bibitem[\protect\citeauthoryear{{Oke} \& {Gunn}}{{Oke} \&
  {Gunn}}{1983}]{oke1983}
{Oke} J.~B.,  {Gunn} J.~E.,  1983, \mn@doi [\apj] {10.1086/160817}, \href
  {https://ui.adsabs.harvard.edu/abs/1983ApJ...266..713O} {266, 713}

\bibitem[\protect\citeauthoryear{{Pontoppidan} et~al.,}{{Pontoppidan}
  et~al.}{2022}]{Pontoppidan2022}
{Pontoppidan} K.,  et~al., 2022, arXiv e-prints, \href
  {https://ui.adsabs.harvard.edu/abs/2022arXiv220713067P} {p. arXiv:2207.13067}

\bibitem[\protect\citeauthoryear{{Roberts-Borsani} et~al.,}{{Roberts-Borsani}
  et~al.}{2022}]{roberts-borsani2022}
{Roberts-Borsani} G.,  et~al., 2022, arXiv e-prints, \href
  {https://ui.adsabs.harvard.edu/abs/2022arXiv221015639R} {p. arXiv:2210.15639}

\bibitem[\protect\citeauthoryear{{Robertson}, {Ellis}, {Dunlop}, {McLure}  \&
  {Stark}}{{Robertson} et~al.}{2010}]{robertson2010}
{Robertson} B.~E.,  {Ellis} R.~S.,  {Dunlop} J.~S.,  {McLure} R.~J.,   {Stark}
  D.~P.,  2010, \mn@doi [\nat] {10.1038/nature09527}, \href
  {https://ui.adsabs.harvard.edu/abs/2010Natur.468...49R} {468, 49}

\bibitem[\protect\citeauthoryear{{Rogers}, {McLure}  \& {Dunlop}}{{Rogers}
  et~al.}{2013}]{rogers2013}
{Rogers} A.~B.,  {McLure} R.~J.,   {Dunlop} J.~S.,  2013, \mn@doi [\mnras]
  {10.1093/mnras/sts515}, \href
  {https://ui.adsabs.harvard.edu/abs/2013MNRAS.429.2456R} {429, 2456}

\bibitem[\protect\citeauthoryear{{Rogers} et~al.,}{{Rogers}
  et~al.}{2014}]{rogers2014}
{Rogers} A.~B.,  et~al., 2014, \mn@doi [\mnras] {10.1093/mnras/stu558}, \href
  {https://ui.adsabs.harvard.edu/abs/2014MNRAS.440.3714R} {440, 3714}

\bibitem[\protect\citeauthoryear{{Schaerer}}{{Schaerer}}{2002}]{schaerer2002}
{Schaerer} D.,  2002, \mn@doi [\aap] {10.1051/0004-6361:20011619}, \href
  {https://ui.adsabs.harvard.edu/abs/2002A&A...382...28S} {382, 28}

\bibitem[\protect\citeauthoryear{{Smith}, {Kannan}, {Garaldi}, {Vogelsberger},
  {Pakmor}, {Springel}  \& {Hernquist}}{{Smith} et~al.}{2022}]{smith2022}
{Smith} A.,  {Kannan} R.,  {Garaldi} E.,  {Vogelsberger} M.,  {Pakmor} R.,
  {Springel} V.,   {Hernquist} L.,  2022, \mn@doi [\mnras]
  {10.1093/mnras/stac713}, \href
  {https://ui.adsabs.harvard.edu/abs/2022MNRAS.512.3243S} {512, 3243}

\bibitem[\protect\citeauthoryear{{Somerville}, {Gilmore}, {Primack}  \&
  {Dom{\'\i}nguez}}{{Somerville} et~al.}{2012}]{somerville2012}
{Somerville} R.~S.,  {Gilmore} R.~C.,  {Primack} J.~R.,   {Dom{\'\i}nguez} A.,
  2012, \mn@doi [\mnras] {10.1111/j.1365-2966.2012.20490.x}, \href
  {https://ui.adsabs.harvard.edu/abs/2012MNRAS.423.1992S} {423, 1992}

\bibitem[\protect\citeauthoryear{{Somerville}, {Popping}  \&
  {Trager}}{{Somerville} et~al.}{2015}]{somerville2015}
{Somerville} R.~S.,  {Popping} G.,   {Trager} S.~C.,  2015, \mn@doi [\mnras]
  {10.1093/mnras/stv1877}, \href
  {https://ui.adsabs.harvard.edu/abs/2015MNRAS.453.4337S} {453, 4337}

\bibitem[\protect\citeauthoryear{{Somerville} et~al.,}{{Somerville}
  et~al.}{2021}]{somerville2021}
{Somerville} R.~S.,  et~al., 2021, \mn@doi [\mnras] {10.1093/mnras/stab231},
  \href {https://ui.adsabs.harvard.edu/abs/2021MNRAS.502.4858S} {502, 4858}

\bibitem[\protect\citeauthoryear{{Speagle}}{{Speagle}}{2020}]{speagle2020}
{Speagle} J.~S.,  2020, \mn@doi [\mnras] {10.1093/mnras/staa278}, \href
  {https://ui.adsabs.harvard.edu/abs/2020MNRAS.493.3132S} {493, 3132}

\bibitem[\protect\citeauthoryear{{Tacchella} et~al.,}{{Tacchella}
  et~al.}{2022}]{tacchella2022}
{Tacchella} S.,  et~al., 2022, \mn@doi [\apj] {10.3847/1538-4357/ac4cad}, \href
  {https://ui.adsabs.harvard.edu/abs/2022ApJ...927..170T} {927, 170}

\bibitem[\protect\citeauthoryear{{Topping}, {Stark}, {Endsley}, {Plat},
  {Whitler}, {Chen}  \& {Charlot}}{{Topping} et~al.}{2022}]{topping2022}
{Topping} M.~W.,  {Stark} D.~P.,  {Endsley} R.,  {Plat} A.,  {Whitler} L.,
  {Chen} Z.,   {Charlot} S.,  2022, arXiv e-prints, \href
  {https://ui.adsabs.harvard.edu/abs/2022arXiv220801610T} {p. arXiv:2208.01610}

\bibitem[\protect\citeauthoryear{{Treu} et~al.,}{{Treu}
  et~al.}{2022}]{treu2022}
{Treu} T.,  et~al., 2022, arXiv e-prints, \href
  {https://ui.adsabs.harvard.edu/abs/2022arXiv220607978T} {p. arXiv:2206.07978}

\bibitem[\protect\citeauthoryear{{V{\'a}zquez}, {Leitherer}, {Heckman},
  {Lennon}, {de Mello}, {Meurer}  \& {Martin}}{{V{\'a}zquez}
  et~al.}{2004}]{vazquez2004}
{V{\'a}zquez} G.~A.,  {Leitherer} C.,  {Heckman} T.~M.,  {Lennon} D.~J.,  {de
  Mello} D.~F.,  {Meurer} G.~R.,   {Martin} C.~L.,  2004, \mn@doi [\apj]
  {10.1086/379805}, \href
  {https://ui.adsabs.harvard.edu/abs/2004ApJ...600..162V} {600, 162}

\bibitem[\protect\citeauthoryear{{Vijayan}, {Lovell}, {Wilkins}, {Thomas},
  {Barnes}, {Irodotou}, {Kuusisto}  \& {Roper}}{{Vijayan}
  et~al.}{2021}]{vijayan2021}
{Vijayan} A.~P.,  {Lovell} C.~C.,  {Wilkins} S.~M.,  {Thomas} P.~A.,  {Barnes}
  D.~J.,  {Irodotou} D.,  {Kuusisto} J.,   {Roper} W.~J.,  2021, \mn@doi
  [\mnras] {10.1093/mnras/staa3715}, \href
  {https://ui.adsabs.harvard.edu/abs/2021MNRAS.501.3289V} {501, 3289}

\bibitem[\protect\citeauthoryear{{Wilkins}, {Bouwens}, {Oesch}, {Labb{\'e}},
  {Sargent}, {Caruana}, {Wardlow}  \& {Clay}}{{Wilkins}
  et~al.}{2016}]{wilkins2016}
{Wilkins} S.~M.,  {Bouwens} R.~J.,  {Oesch} P.~A.,  {Labb{\'e}} I.,  {Sargent}
  M.,  {Caruana} J.,  {Wardlow} J.,   {Clay} S.,  2016, \mn@doi [\mnras]
  {10.1093/mnras/stv2263}, \href
  {https://ui.adsabs.harvard.edu/abs/2016MNRAS.455..659W} {455, 659}

\bibitem[\protect\citeauthoryear{{Yung}, {Somerville}, {Finkelstein}, {Popping}
   \& {Dav{\'e}}}{{Yung} et~al.}{2019a}]{yung2019a}
{Yung} L.~Y.~A.,  {Somerville} R.~S.,  {Finkelstein} S.~L.,  {Popping} G.,
  {Dav{\'e}} R.,  2019a, \mn@doi [\mnras] {10.1093/mnras/sty3241}, \href
  {https://ui.adsabs.harvard.edu/abs/2019MNRAS.483.2983Y} {483, 2983}

\bibitem[\protect\citeauthoryear{{Yung}, {Somerville}, {Popping},
  {Finkelstein}, {Ferguson}  \& {Dav{\'e}}}{{Yung} et~al.}{2019b}]{yung2019b}
{Yung} L.~Y.~A.,  {Somerville} R.~S.,  {Popping} G.,  {Finkelstein} S.~L.,
  {Ferguson} H.~C.,   {Dav{\'e}} R.,  2019b, \mn@doi [\mnras]
  {10.1093/mnras/stz2755}, \href
  {https://ui.adsabs.harvard.edu/abs/2019MNRAS.490.2855Y} {490, 2855}

\bibitem[\protect\citeauthoryear{{Yung} et~al.,}{{Yung}
  et~al.}{2022a}]{yung2022b}
{Yung} L.~Y.~A.,  et~al., 2022a, arXiv e-prints, \href
  {https://ui.adsabs.harvard.edu/abs/2022arXiv221004902Y} {p. arXiv:2210.04902}

\bibitem[\protect\citeauthoryear{{Yung} et~al.,}{{Yung}
  et~al.}{2022b}]{yung2022}
{Yung} L.~Y.~A.,  et~al., 2022b, \mn@doi [\mnras] {10.1093/mnras/stac2139},
  \href {https://ui.adsabs.harvard.edu/abs/2022MNRAS.515.5416Y} {515, 5416}

\bibitem[\protect\citeauthoryear{{Zheng} et~al.,}{{Zheng}
  et~al.}{2012}]{zheng2012}
{Zheng} W.,  et~al., 2012, \mn@doi [\nat] {10.1038/nature11446}, \href
  {https://ui.adsabs.harvard.edu/abs/2012Natur.489..406Z} {489, 406}

\makeatother
\end{thebibliography}





\bsp	
\label{lastpage}
\end{document}